\documentclass[letterpaper,twocolumn,10pt]{article}
\usepackage{usenix-2020-09}
\pdfoutput=1
\usepackage{tikz}
\usepackage{amsmath}
\usepackage{url}

\usepackage{graphicx}
\usepackage{tabularx}
\usepackage{color}
\usepackage{array}
\usepackage{makecell}
\usepackage{multirow}
\usepackage{enumerate}
\usepackage{blindtext}
\usepackage{caption}
\usepackage{subfigure}
\usepackage{booktabs}
\usepackage{tcolorbox}
\usepackage{amsmath}
\usepackage{cleveref}
\usepackage{breakurl}
\usepackage{diagbox}
\usepackage{balance}
\usepackage{vcell}
\usepackage{algorithm}
\usepackage{algorithmic}
\usepackage{balance}

\begin{document}

\date{}

\title{Exploring Unconfirmed Transactions for Effective Bitcoin Address Clustering}



\author{
{\rm Kai Wang}\\
Fudan University
\and
{\rm Maike Tong}\\
Fudan University
\and
{\rm Changhao Wu}\\
Fudan University
\and
{\rm Jun Pang}\\
University of Luxembourg
\and
{\rm Chen Chen}\\
Fudan University
\and
{\rm Xiapu Luo}\\
The Hong Kong Polytechnic University
\and
{\rm Weili Han}\\
Fudan University
} 

\maketitle

\begin{abstract}
The development of \emph{clustering heuristics} has demonstrated that Bitcoin is not completely anonymous. Currently, existing clustering heuristics only consider \emph{confirmed} transactions recorded in the Bitcoin blockchain. However, \emph{unconfirmed} transactions in the mempool have yet to be utilized to improve the performance of the clustering heuristics.

In this paper, we bridge this gap by combining unconfirmed and confirmed transactions for clustering Bitcoin addresses effectively. First, we present a data collection system for capturing unconfirmed transactions. Two case studies are performed to show the presence of user behaviors in unconfirmed transactions not present in confirmed transactions. Next, we apply the state-of-the-art clustering heuristics to unconfirmed transactions, and the clustering results can reduce the number of entities after applying, for example, the co-spend heuristics in confirmed transactions by 2.3\%. Finally, we propose three novel clustering heuristics to capture specific behavior patterns in unconfirmed transactions, which further reduce the number of entities after the application of the co-spend heuristics by 9.8\%. Our results demonstrate the utility of unconfirmed transactions in address clustering and further shed light on the limitations of anonymity in cryptocurrencies. To the best of our knowledge, this paper is the first to apply the unconfirmed transactions in Bitcoin to cluster addresses.

\end{abstract}


\section{Introduction}

Since its introduction in 2008, Bitcoin~\cite{BTCWhitePaper} has aimed to provide an anonymous payment system that can prevent user privacy from leakages. In Bitcoin, users initiate transactions where they and their recipients use a set of addresses as pseudonyms to represent them. Bitcoin users can generate any number of addresses with no inherent link to their personal identities. In other words, every Bitcoin user can anonymize transactions using a new address (i.e., one-time address) for each newly initiated transaction. These one-time addresses can break the association among addresses controlled by the same Bitcoin user, thereby protecting Bitcoin user privacy. Attracted by the above design, an increasing number of users conduct Bitcoin transactions to carry out commercial activities~\cite{DBLP:conf/socialcom/ReidH11}. As of December 31, 2022, there are around 800 million transactions in Bitcoin. 


\begin{figure*}[htbp]
    \centering
    \includegraphics[scale=0.24]{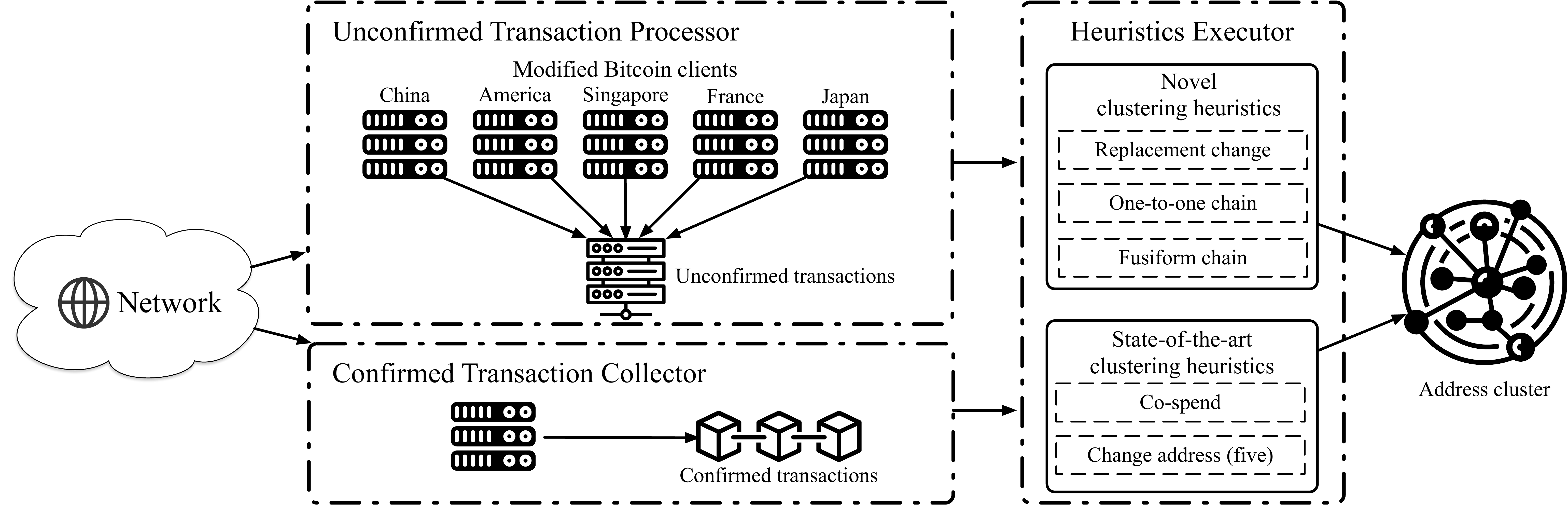}
    \caption{A new procedure for clustering Bitcoin addresses by combining unconfirmed transactions and confirmed transactions.}
    \label{fig:overview}
\end{figure*}


Bitcoin's inability to attain complete anonymity has been confirmed despite utilizing addresses as pseudonyms. The emergence of various \emph{clustering heuristics} used to detect multiple addresses controlled by the same entity has been the primary driving force behind this outcome~\cite{DBLP:conf/imc/MeiklejohnPJLMVS13, BitcoinTool, DBLP:journals/corr/abs-1709-02489, Reid2012An, DBLP:journals/tweb/WangPCZHCH22, DBLP:conf/uss/KapposYSRHM22}. Transactions with multiple inputs usually arise if the user does not have a single UTXO large enough to fulfill the payment amount. As mentioned in the studies~\cite{DBLP:conf/imc/MeiklejohnPJLMVS13,BitcoinTool,DBLP:journals/corr/abs-1709-02489}, one direct idea, called \emph{co-spend}, assumes that all input addresses used for one Bitcoin transaction should belong to the same entity. This clustering heuristic is widely used in practice by blockchain analytics companies like Chainalysis~\cite{Chainalysis}. As it is difficult to send the exact specified amount of funds to the receiver, the remainder of the funds are to be returned to the sender via a change address. Therefore, the change address heuristic is designed to identify the address in the transaction outputs used to receive the change~\cite{DBLP:journals/corr/abs-1709-02489}. The latest work~\cite{DBLP:conf/uss/KapposYSRHM22} tracks a particular type of funds flow, called a \emph{peel chain}, which represents many transactions carried out by the same entity. Over the years, a number of \emph{clustering heuristics} have been proposed, refined, and widely used, helping the analysis of Bitcoin transactions shift from the address perspective to the user perspective, which has significantly improved the understanding of the Bitcoin ecosystem.


So far, existing clustering heuristics focus only on \emph{confirmed} transactions but ignore \emph{unconfirmed} transactions not yet recorded in the blockchain. When a user initiates a transaction, it is verified and propagated across the Bitcoin network. The node that receives this transaction stores it in a temporary storage area known as the \emph{mempool}. Miners select transactions from their respective mempools to form a block and record these transactions in the blockchain. These transactions recorded in the blockchain are called \emph{confirmed transactions}, while those not recorded in the blockchain are called \emph{unconfirmed transactions}. There is no guarantee whether and when a transaction will be recorded in the blockchain, as some transactions may be replaced or have insufficient transaction fees. All along, researchers have focused only on confirmed transactions. However, unconfirmed transactions are equally important for clustering Bitcoin addresses and breaking the user pseudo-anonymity. Unconfirmed transactions are user-initiated and reflect users' transaction intentions, behaviors, and bitcoin usage habits. Moreover, unconfirmed transactions contain additional information about the state of these transactions when they are still in the mempool, which cannot be obtained from confirmed transactions. Therefore, analyzing the details of unconfirmed transactions and the relationship between unconfirmed transactions and confirmed transactions can play a significant role in Bitcoin address clustering.

To the best of our knowledge, we are the first to cluster Bitcoin addresses by combining unconfirmed and confirmed transactions. Figure~\ref{fig:overview} illustrates the optimized process for clustering Bitcoin addresses. First, we establish a data collection system consisting of the Confirmed Transaction Collector~(CTC) and the Unconfirmed Transaction Processor~(UTP). The CTC with one Bitcoin client is responsible for recording confirmed transactions in the blockchain. The UTP with five modified Bitcoin clients is tasked with recording unconfirmed transactions in real time. Besides, to ensure the UTP can record as many as possible transactions broadcast on the Bitcoin network, we deliberately establish these five clients in five countries with which we have a constant and stable network connection. To demonstrate the importance of unconfirmed transactions, we conduct two exploratory case studies and apply the state-of-the-art clustering heuristics to unconfirmed transactions in the Heuristic Executor. Our findings reveal that unconfirmed transactions have a significant impact on the results of these heuristics, with the co-spend result being the most affected one -- its clustering result is reduced by 1,011,406 entities, representing 2.3\% of the total number of entities. 

Next, our subsequent investigation into the particular transaction behaviors of users in unconfirmed transactions reveals further associations between addresses that are beyond the scope of the existing heuristics. Based on these findings, we design three novel clustering heuristics for the Heuristic Executor by analyzing the relationship between unconfirmed and confirmed transactions through the mechanisms of Bitcoin mempool, such as the Replace-by-fee~(RBF) proposed by BIP125. These novel clustering heuristics have an even more profound impact on each of the clustering results, with the most affected being the co-spend clustering result where the number of entities is further reduced by 4,126,352, representing 9.8\% of the total number of entities.



In summary, we are motivated to cluster Bitcoin addresses by combining unconfirmed and confirmed transactions, to improve the state-of-the-art clustering heuristics:
\begin{itemize}
    \item We establish a data collection system to record transactions received in real-time. We release the dataset\footnote{\url{https://drive.google.com/drive/folders/1Vc5p9qro8zh6lV6lLqQMB4AtSvdhLjiT?usp=sharing}} as a benchmark for future studies. Moreover, we apply the state-of-the-art clustering heuristics to unconfirmed transactions. Unconfirmed transactions affect the results of each of these clustering heuristics, with the co-spend clustering result being the most affected -- the number of entities is reduced by 2.3\%.
    \item We design three novel clustering heuristics by analyzing the particular transaction behavior of users in unconfirmed transactions, to further expand the clustering result. The most affected is the co-spend clustering result where the number of entities is reduced by 4,126,352, representing 9.8\% of the total number of entities. To the best of our knowledge, this paper is the first to explore unconfirmed transactions to cluster addresses in Bitcoin.
\end{itemize}



\smallskip\noindent
\textbf{Roadmap of the paper}. 
Section~\ref{sec:background} provides a brief introduction to Bitcoin address clustering and Bitcoin mempool. We collect unconfirmed transactions, present two exploratory case studies on unconfirmed transactions, and apply the state-of-the-art clustering heuristics to unconfirmed transactions to expand the clustering result of confirmed transactions in Section~\ref{sec:analysis}. In Section~\ref{sec:novel}, we design three novel clustering heuristics that further improve the clustering results. We discuss the achievements and associated limitations of our method in Section~\ref{sec:discussion}. Finally, we review related work in Section~\ref{sec:relateworks}, and conclude this paper in Section~\ref{sec:conclusion}.


%

\section{Background}
\label{sec:background}

\subsection{Bitcoin Address Clustering}
Bitcoin aims to provide an anonymous payment system that allows users to remain anonymous with the ownership of their addresses even if all transactions are publicly available, thereby completing transactions reliably and minimizing privacy exposure. First, transactions are publicly available to ensure their integrity. Bitcoin then records transactions as a block, and adds it into an extended proof-of-work chain. Second, to conceal users' real-world identities, Bitcoin provides addresses as pseudonyms in transactions. It is recommended that users generate new addresses for each transaction to break the association between Bitcoin addresses controlled by the same user. Here, a Bitcoin address is a hash string of a public key, consisting of digits and characters, and can be generated at any time without requiring identity verification. This practice helps to break the link between owners and their addresses, thereby concealing the ownership of the Bitcoin addresses.

The development of various \emph{clustering heuristics} has demonstrated that Bitcoin is not completely anonymous. The \emph{clustering heuristics} rely on the transactional behavior observed in the Bitcoin blockchain to indicate that the addresses they cluster together are controlled by the same entity. A valid Bitcoin transaction must be signed using the private keys associated with all inputs of this transaction. This mechanism has given rise to a common heuristic that assumes all inputs of a transaction are controlled by the same entity, known as the \emph{co-spend} or \emph{multi-input} heuristic. Although this heuristic is extensively validated and widely used in practice, the emergence of \emph{Coinjoin} transactions causes this heuristic to fail. \emph{CoinJoin} is a trustless method for combining multiple Bitcoin payments from multiple spenders into a single transaction to make it more difficult to determine which sender paid which recipient or recipients. Therefore, several works filter out \emph{Coinjoin} transactions before applying this heuristic.

Bitcoin uses the Unspent Transaction Output~(UTXO) to represent value. A UTXO represents a certain amount of bitcoins that must be used as a complete unit. This results in the need for senders to use a change address to receive the remaining amount of bitcoins. Determining the change address of senders has become the focus of academia and industry. There are a number of proposed \emph{change address} heuristics~\cite{DBLP:conf/fc/AndroulakiKRSC13, DBLP:conf/imc/MeiklejohnPJLMVS13, DBLP:journals/popets/GoldfederKRN18, DBLP:conf/icmla/ErmilovPY17, DBLP:conf/uss/KapposYSRHM22} to identify the change address in a Bitcoin transaction.

Existing clustering heuristics focus only on transaction patterns presented by transactions already recorded in the Bitcoin blockchain. These clustering heuristics ignore the information of a transaction \emph{before} it is recorded in the blockchain.

\subsection{Bitcoin Mempool}
Bitcoin mempool~(a contraction of memory and pool) is a mechanism for temporarily storing information about unconfirmed transactions. It serves as a sort of waiting room for transactions that have been validated but have not yet been included in a block. Each Bitcoin node maintains its own mempool for storing transactions received and validated by this node.

\begin{figure}
    \centering
    \includegraphics[scale=0.35]{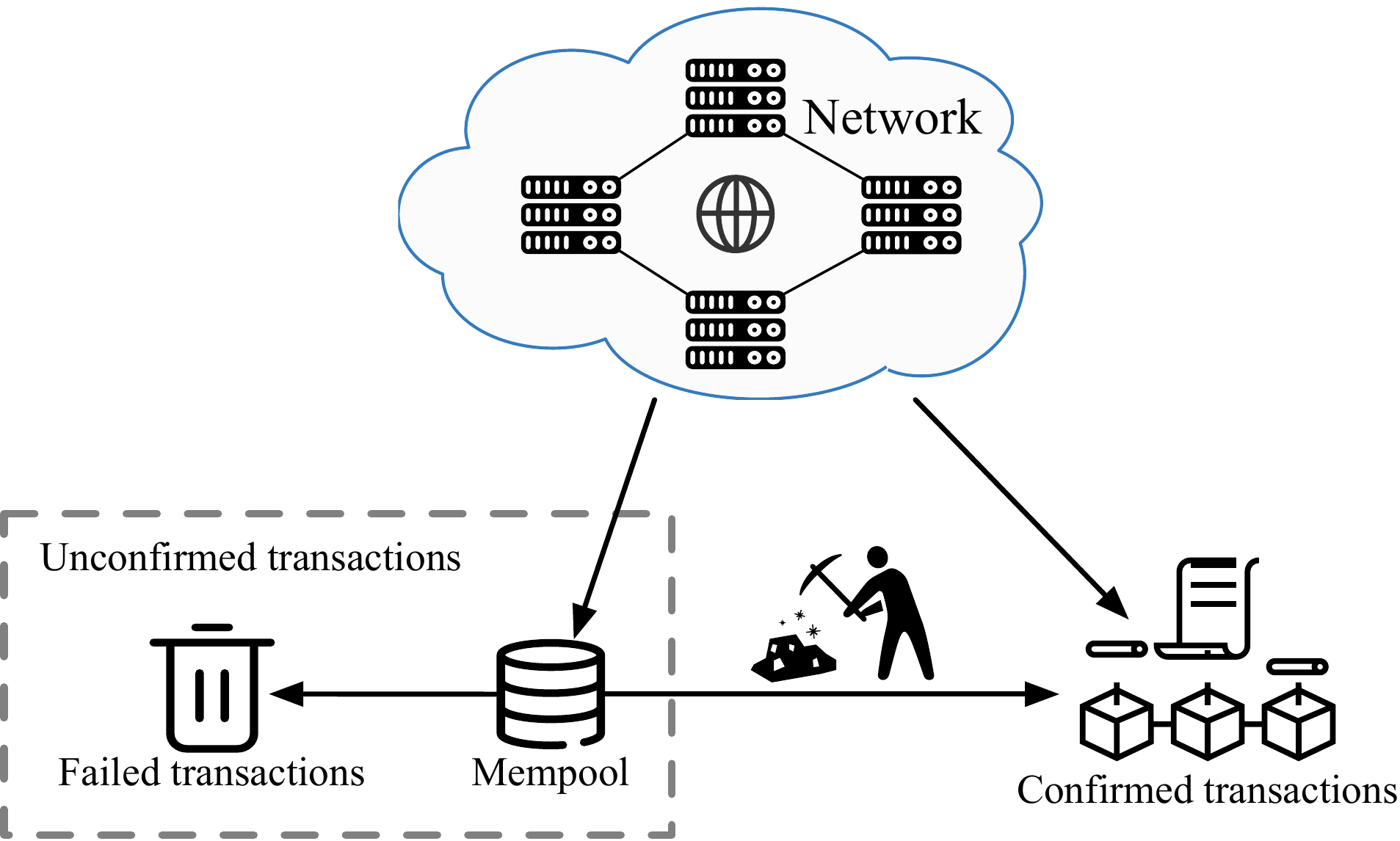}
    \caption{Life-cycle of a Bitcoin transaction.}
    \label{fig:Life-cycle}
\end{figure}

Once a Bitcoin transaction is propagated to any node connected to the bitcoin network, the node will validate (or reject) the transaction according to a set criterion. For example, it will ensure that the signature is correct and the amount of the outputs doesn’t exceed that of the inputs. Then, if the transaction is found to be valid, the node will propagate the transaction to other connected nodes and add it to its mempool. As shown in Figure~\ref{fig:Life-cycle}, all Bitcoin transactions just initiated by users are unconfirmed transactions and only exist in the mempool until they’re recorded in the blockchain. Miners are responsible for selecting transactions from the mempool and recording them in the blockchain, where these transactions become confirmed transactions. Some unconfirmed transactions may never be confirmed due to double-spending or other issues, and these transactions can be considered failed transactions, while others remain in the mempool waiting for confirmation. In summary, unconfirmed transactions consist of two parts: failed transactions and transactions that are currently in the mempool waiting to be selected.

We often speak of the mempool, but it should be noted that there is no universal mempool shared by all nodes. Each Bitcoin mempool is configured differently and stores different transactions. Lower-end devices with limited resources may only dedicate small amounts of memory to recording transactions, whereas higher-end ones might devote considerably more memory.

Every unconfirmed transaction in the mempool has four specific fields named \emph{bip125-replaceable}, \emph{time}, \emph{depends}, and \emph{spentby}. These fields are available only when a transaction is in Bitcoin mempool, and will not exist after the transaction is recorded in the blockchain. The \emph{bip125-replaceable} is a Boolean value that indicates whether this transaction can be replaced through another transaction with a higher fee. The \emph{time} represents the time when the transaction enters the mempool of the current node and the \emph{time} usually has slight differences in different mempools. The \emph{depends} of a transaction records unconfirmed transactions used as inputs for this transaction. The \emph{spentby} records unconfirmed transactions spending outputs of this transaction. These above fields contain rich information about a transaction before it is confirmed, which can be utilized for Bitcoin address clustering. The transaction data in the mempool changes in real time, as unconfirmed transactions are constantly added to the mempool, and confirmed or failed transactions are subsequently removed from the mempool. 


The analysis of unconfirmed transactions in the mempool is still in its infancy. Therefore, the available research results are relatively few.

\section{Dataset and Analysis}
\label{sec:analysis}

In this section, we provide details of the Unconfirmed Transaction (UTP), which captures and processes unconfirmed transactions, as illustrated in Figure~\ref{fig:overview}. Furthermore, we demonstrate the value of unconfirmed transactions through two exploratory case studies and by applying the state-of-the-art clustering heuristics to unconfirmed transactions.

\subsection{Data Collector and Processor}
The Confirmed Transaction Collector~(CTC) with one Bitcoin client is to synchronize confirmed transactions in the blockchain. Meanwhile, the UTP with 5 modified Bitcoin clients is to record all unconfirmed transactions broadcast on the Bitcoin network and continuously reconstruct the state of Bitcoin mempool. This state comprises information on the number of transactions in the mempool and the details of each transaction.

Initially, we modify the Bitcoin client \emph{bitcoind} to capture unconfirmed transactions in the mempool where transactions have passed an initial check, such as ensuring the total amount of transaction outputs is greater than the total amount of transaction inputs. Since each mempool is set up differently and receives transactions at varying times, we established 5 modified Bitcoin clients in China, America, Singapore, France, and Japan to capture as many unconfirmed transactions as possible broadcast on the Bitcoin network. Table~\ref{table:extra_details} shows the extra information of an unconfirmed transaction in the mempool compared to a confirmed transaction in the blockchain. In this paper, we focus on 6 fields, including \emph{fee}, \emph{vsize}, \emph{time}, \emph{removetime}, \emph{depends}, \emph{spentby}, and \emph{replaceable}.

Subsequently, we build a mempool state database based on TiDB~\cite{PingCAP}. In this database, we set \emph{time} and \emph{removetime} as indexes for each unconfirmed transaction. Given a specific time, the database is able to retrieve all unconfirmed transactions in the mempool at that moment according to the \emph{time} and \emph{removetime}.

\begin{table}
\centering
\caption{Other details of an unconfirmed transaction in the mempool.}
\scalebox{0.80}{
\begin{tabular}{ll} 
\toprule
\textbf{Name}   & \textbf{Content}                                                                         \\ 
\hline
fee            & transaction fee in BTC.                                                                 \\
vsize           & virtual transaction size.                                                               \\
weight          & transaction weight.                                                                      \\
time            & local time when transaction entered pool.                                                \\
removetime      & local time when transaction are removed.                                                 \\
height          & block height transaction entered pool.                                                   \\
descendantcount & \begin{tabular}[c]{@{}l@{}}number of in-mempool \\descendant transactions.\end{tabular}  \\
descendantsize  & vsize of in-mempool descendants.                                                         \\
descendantfees  & modified fees of in-mempool descendants.                                                 \\
ancestorcount   & \begin{tabular}[c]{@{}l@{}}number of in-mempool \\ancestor transactions.\end{tabular}    \\
ancestorsize    & vsize of in-mempool ancestors.                                                           \\
ancestorfees    & modified fees of in-mempool ancestors.                                                   \\
wtxid           & hash of transaction, including witness data.                                             \\
depends         & unconfirmed transactions used as inputs.                                                 \\
spentby         & unconfirmed transactions spending outputs.                                               \\
replaceable     & whether this transaction could be replaced.          \\
\bottomrule
\end{tabular}
}
\label{table:extra_details}
\end{table}

\begin{figure*}[htbp]
    \centering
    \includegraphics[scale=0.52]{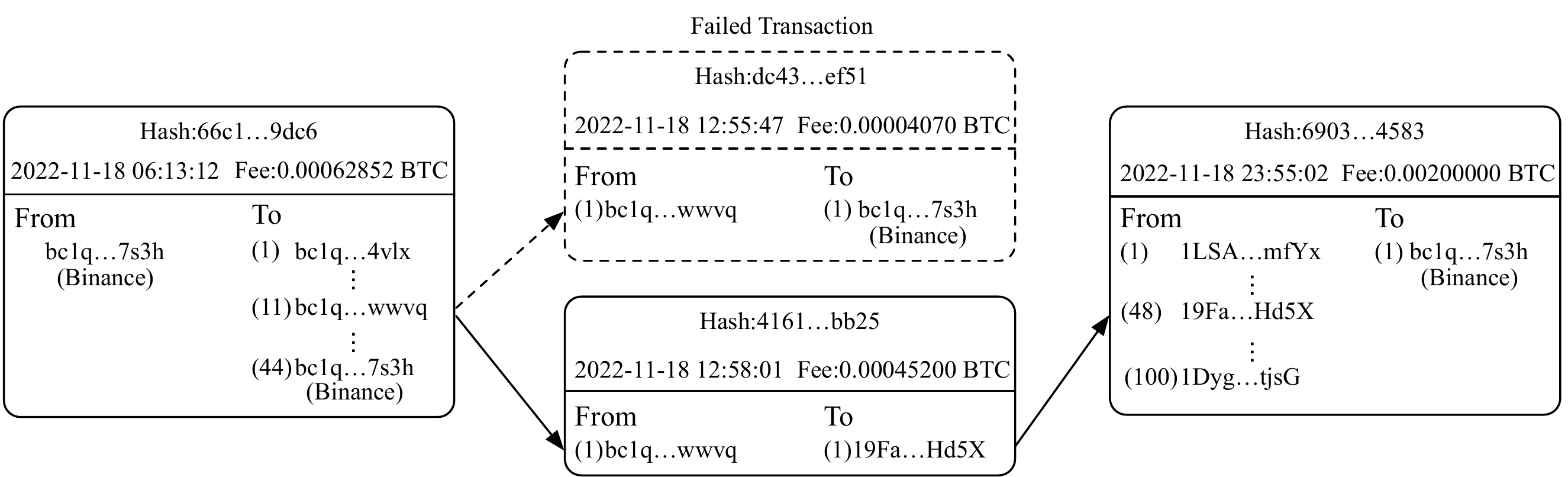}
    \caption{Binance exchange address analysis.}
    \label{fig:Binance}
\end{figure*}

Since there are transaction failures in unconfirmed transactions, we make specific treatments to the transaction structure. First, the original transaction input contains only UTXO and no address information. We locate the address information corresponding to this UTXO based on the previous transaction hash and transaction output number included in the UTXO, and then add the address information to the transaction input. Second, a UTXO may be used by multiple unconfirmed transactions as their inputs. Eventually, only one of these transactions gets confirmed, while the rest become failed transactions. Therefore, we design two fields in the transaction output to represent the information related to its being spent. Specifically, the \emph{output.is\_spent} field, a Boolean type, indicates whether the \emph{output} has been used as an input in other transactions. The \emph{output.spent\_tx} field is a list containing transactions that use the \emph{output} as their inputs.


We start collecting data on May 1, 2022 and collect a total of 51,216,932 transactions by December 31, 2022. Of these, 1,415,140 transactions fail to be confirmed and become failed transactions. These failed transactions involve 6,475,103 Bitcoin addresses, with 601,553 of them not involved in any confirmed transactions. These addresses have been generated by users but have yet to be recorded in the blockchain. As a result, no information about these addresses is visible in the Bitcoin blockchain. However, these addresses could be used by users to participate in transactions in the future, so analyzing them could provide early warning information.


\subsection{Case Studies}
Although unconfirmed transactions or even failed transactions are not recorded in the blockchain, they still reveal the motivations behind why users initiate these transactions. This information is highly valuable for analyzing user behaviors. In this section, we demonstrate the utility of failed transactions in analyzing user behaviors through two case studies.

\subsubsection{Binance exchange address analysis}

One of the main safeguards used by Bitcoin exchanges to prevent attacks is the use of cold and hot storage technology within the exchange, with carefully designed risk control systems~\cite{DBLP:conf/blocksys/LiLZ19}. There are three main types of addresses controlled by Bitcoin exchanges: hot wallet addresses, cold wallet addresses, and user wallet addresses. Therefore, there is a clear division of labor at the addresses of a Bitcoin exchange.

Hot wallets are always connected to the Internet. The main function of the hot wallet is to maintain a bitcoin pool for the user’s withdrawal demand. The user wallet address is created by the Bitcoin exchange and the private key of the address is held by the Bitcoin exchange. Users can deposit bitcoins to Bitcoin exchange by the user wallet address.

Figure~\ref{fig:Binance} shows a failed transaction associated with the Binance exchange, a confirmed transaction to replace the failed transaction, and another two confirmed transactions related to Binance exchange. The address \emph{bc1q...7s3h} is labeled with the Binance exchange by Blockchain.com~\cite{blockchain.com}. When the transaction \emph{66c1...9dc6} is confirmed, the transaction \emph{dc43...ef51} is initiated to transfer bitcoins in the address \emph{bc1q...wwvq} to Binance exchange address. Three minutes after the transaction \emph{dc43...ef51} is initiated, the same user initiates the transaction \emph{4161... bb25}. The second transaction spends the same UTXO as the first transaction but has a much higher fee. As a result, miners choose the second transaction to pack into a block that is eventually confirmed. The transaction \emph{dc43...ef51} turns out to be a failed transaction that is removed by all Bitcoin nodes and can never be confirmed again. Then, another transaction \emph{6903...4583} is initiated with 100 inputs and 1 output, which transfers bitcoins in the address \emph{19Fa...Hd5X} to the address \emph{bc1q...7s3h}.

In this scenario, the transaction \emph{dc43...ef51} and the transaction \emph{4161... bb25} are initiated by the same user because they spend the same UTXO in the address \emph{bc1q...wwvq}. From the content of the transaction \emph{dc43...ef51} and transaction \emph{6903...4583}, the purpose of the address \emph{bc1q...wwvq} is to transfer its bitcoins to the address \emph{bc1q...7s3h}. A question worth analyzing is why the user would replace the transaction \emph{dc43...ef51} with the second one \emph{4161...bb25}.

To answer this question, we investigate 127 confirmed transactions in which the address \emph{bc1q...wwvq} is involved. All outgoing transactions of the address \emph{bc1q...wwvq} are consist of one input and one output. When the address \emph{bc1q...wwvq} transfers bitcoins to other addresses, the recipient address is always the address \emph{19Fa...Hd5X}. Then, the address \emph{19Fa...Hd5X} transfer bitcoins to the address \emph{bc1q...7s3h}. Both incoming and outgoing transactions of the address \emph{bc1q...wwvq} occur in pairs, i.e. the address receives bitcoins and then transfers out all the bitcoins it receives. Therefore, we infer that the transaction \emph{dc43...ef5} was a mistake by the Binance exchange when operating the bitcoin transfer. Address \emph{bc1q...7s3h} is a hot wallet address of Binance exchange, and under the rules of Binance exchange, address \emph{bc1q} cannot transfer bitcoins directly to address \emph{bc1q...7s3h}. Therefore, when the transaction \emph{dc43...ef51} was initiated, the operator or script discovered the misoperation and in turn initiated another high-fee transaction \emph{4161...bb25} to replace the first one.

As can be seen, the addresses controlled by the Binance exchange have distinct roles and perform specific responsibilities. The Binance exchange carefully plans the transaction relationships between the addresses to prevent the casual transfer of bitcoins between them.

Through analysis of unconfirmed transactions, we can shed light on the Binance exchange's internal risk prevention and control mechanisms, and can assist regulators in verifying the cryptocurrency exchange's reported information and inadvertently disclosed transfer behavior.

\subsubsection{Potential dust attacks against whale addresses}

The dust attack can be defined as malicious behavior targeting Bitcoin users and privacy by sending tiny amounts of bitcoins to victims' addresses~\cite{wang2018anti}. The aim of the dust attacker is to reveal the user's identity by gathering information on where these tiny amounts are combined when the user initiates a new transaction through their cryptocurrency wallet software. The attackers track the transaction activity of these addresses in an attempt to link the dusted addresses and identify the person or company behind them~\cite{wang2018anti}.

\begin{figure}[t]
    \centering
    \includegraphics[scale=0.47]{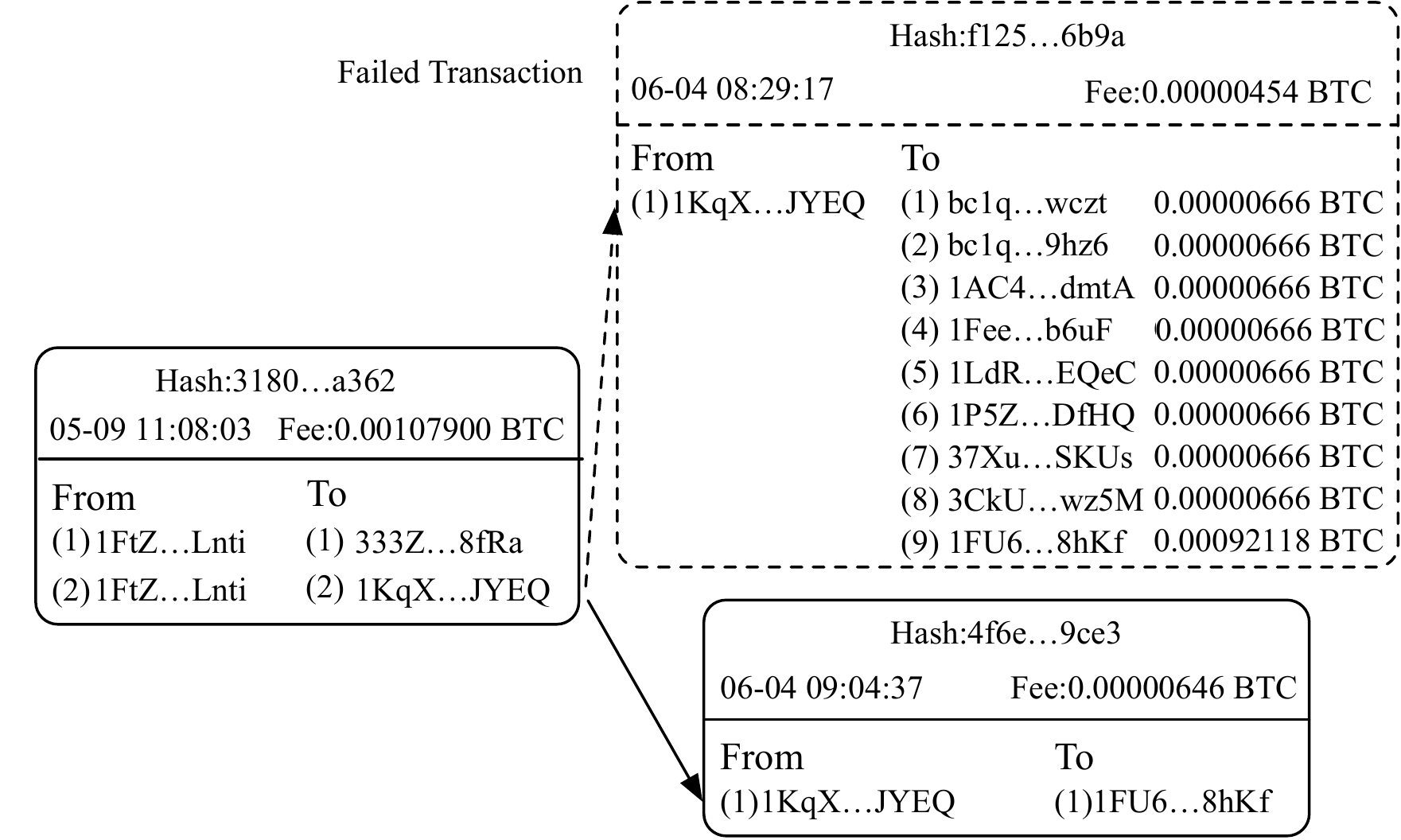}
    \caption{Potential dust attack against whale addresses. }
    \label{fig:Dust}
\end{figure}

Figure~\ref{fig:Dust} shows a potential dust attack against whale addresses found in a failed transaction. Figure~\ref{fig:Dust} contains two confirmed transactions and one failed transaction. In this scenario, the transaction \emph{3180...a362} is first initiated with two inputs and two outputs. When this transaction is confirmed, the address \emph{1KqX...JYEQ} transfers its received bitcoins to the address \emph{1FU6...8hKf} through the transaction \emph{4f6e...9ce3}. We cannot find anything unusual about this address \emph{1KqX...JYEQ} from confirmed transactions.

However, before the transaction \emph{4f6e...9ce3} is initiated, the transaction \emph{f125...6b9a} is first initiated. These two transactions spend the same UTXO \emph{1KqX...JYEQ}. Due to the much higher fee of the transaction \emph{4f6e...9ce3}, miners choose the transaction \emph{4f6e...9ce3} to package into a block that eventually is confirmed. Therefore, the transaction \emph{f125...6b9a} turns out to be a failed transaction. However, the failed transaction \emph{f125...6b9a} reflects the malicious behavior of the user that is not presented in confirmed transactions.

More specifically, the transaction \emph{f125...6b9a} has one input and 9 outputs. It is noteworthy that all 8 outputs of this transaction have the same revenue, i.e. 0.00000666 BTC~(\$ 0.12). The remaining output is the same as the sender and is the change address of the sender. The address \emph{bc1q...wczt}, \emph{bc1q...9hz6} and \emph{1Fee...b6uF} are labeled with \emph{FBI3}, \emph{FBI 2~(Silk Road)} and \emph{MtGox Hacker} respectively by Blockchain.com~\cite{blockchain.com}. The stolen bitcoins of the Bitfinex Hack 2016 continue to converge to the address \emph{bc1q...wczt} that currently has a balance of more than 94,643 BTC without any transfers out. Nearly 70,000 BTC confiscated by the U.S. government from the black market site \emph{Silk Road} are transferred to the address \emph{bc1q...9hz6} in early November, 2020. As of this writing, the bitcoins have not been moved or liquidated. Former Mt.Gox CEO Mark Karpeles confirmed that the bitcoins residing at the address \emph{1Fee...b6uF} were stolen from the Mt.Gox exchange. The address \emph{1P5Z...DfHQ} is controlled by the FTX exchange which declared bankruptcy on November 14, 2022. The other addresses except \emph{1FU6...8hKf} also have a large balance when this failed transaction \emph{f125...6b9a} is initiated. Currently, the balance of the address \emph{3CkU...wz5M} is 0. But, the balance of the address \emph{3CkU...wz5} is more than 50,620 BTC when the failed transaction \emph{f125...6b9a} is initiated. Therefore, the address \emph{1KqX...JYEQ} sends a very small amount of bitcoins to multiple whale addresses via the transaction \emph{f125...6b9a}, with the purpose of a dust attack.

Focusing solely on confirmed transactions in the blockchain fails to uncover hidden users' intentions. On the one hand, unconfirmed transactions help us mine user privacy that confirmed transactions cannot reveal, i.e. transaction intentions. Unconfirmed transactions can also be used to identify malicious behavior that cannot be identified by confirmed transactions alone. On the other hand, unconfirmed transactions allow for the identification of malicious behavior before it takes place.

\subsection{Evaluation of Unconfirmed Transactions}

In order to show the value of unconfirmed transactions in address clustering, we apply the state-of-the-art clustering heuristics to unconfirmed transactions to gain additional association information between addresses.
When no clustering heuristic is applied, we consider each address in the confirmed transactions between May 1, 2022 and December 31, 2022 as an entity. There are a total of 107,607,504 addresses, or 107,607,504 entities, in this time period.
In addition to the co-spend heuristic, we also apply five change address heuristics.

\smallskip\noindent
\textbf{Co-spend}~(short for CS) considers all inputs of a transaction are controlled by the same entity if the transaction is not a Coinjoin transaction. We use the mixing transaction identification algorithm developed by Goldfeder et al.~\cite{DBLP:journals/popets/GoldfederKRN18} that determines whether a transaction is a Coinjoin transaction or not.

\smallskip\noindent
\textbf{Androulaki et al.}~(short for A)~\cite{DBLP:conf/fc/AndroulakiKRSC13} identify the change address of a transaction sender if (1) the transaction must have precisely two outputs; and (2) the address is the only fresh address in the outputs, meaning that it has not been previously used in the blockchain.


\smallskip\noindent
\textbf{Meiklejohn et al.}~(short for M)~\cite{DBLP:conf/imc/MeiklejohnPJLMVS13} identify the change address of a transaction sender if (1) the transaction is not a coinbase transaction; (2) the address is the only fresh address in the outputs; and (3) there are no same addresses in transaction input and transaction output.

\smallskip\noindent
\textbf{Goldfeder et al.}~(short for G)~\cite{DBLP:journals/popets/GoldfederKRN18} utilize the criteria established by Meiklejohn et al.~\cite{DBLP:conf/imc/MeiklejohnPJLMVS13}, but they also add a further condition: (4) the transaction cannot be a Coinjoin transaction.

\smallskip\noindent
\textbf{Ermilov et al.}~(short for E)~\cite{DBLP:conf/icmla/ErmilovPY17} are the pioneers in taking into account not just the behavior of addresses, but also the value they receive in a transaction. To locate the change address in a transaction, they employ the following criteria: (1) the transaction has more or less than two inputs; (2) the transaction has precisely two outputs; (3) there are no same addresses in transaction input and transaction output.; (4) the address is the only fresh address in the outputs; and (5) the value received by this address is precise up to the fourth decimal place.


\noindent
\textbf{Kappos et al.}~(short for K)~\cite{DBLP:conf/uss/KapposYSRHM22} identify the change address in a transaction if (1) the transaction is one node in a \emph{peel chain}; (2) the output is spent and the spent transaction is also a node in this \emph{peel chain}.


To make the description clearer, we define names for each group of clustering results, consisting of two parts: the method and the data. For example, the SC clustering result refers to the result of applying one of the state-of-the-art clustering heuristics to confirmed transactions. The SU clustering result refers to the result of applying one of the state-of-the-art clustering heuristics to unconfirmed transactions. And, we represent the merging operation between clustering results as an additive notation. For example, the SC+SU clustering result refers to the result of merging the SC clustering result and the SU clustering result under the same clustering heuristic.

\begin{figure}[htbp]
    \centering
    \includegraphics[scale=0.55]{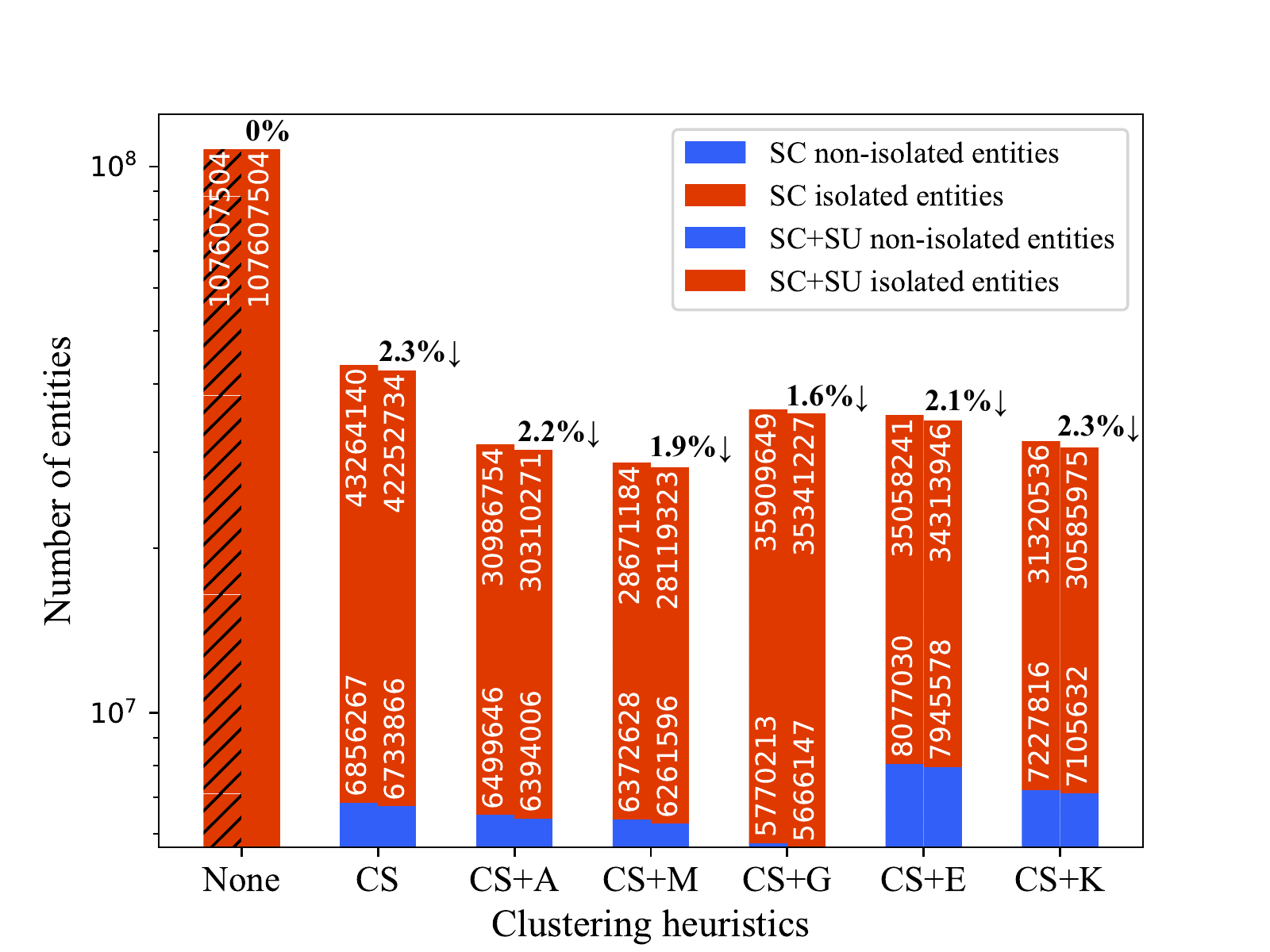}
    \caption{Number of entities in the SC clustering result and the SC+SU clustering result. While an isolated entity has only one Bitcoin address, a non-isolated entity has multiple Bitcoin addresses.}
    \label{fig:compare_1}
\end{figure}

Figure~\ref{fig:compare_1} shows that there is a marked decline in the number of entities in the SC+SU clustering result compared to the SC clustering result in various clustering heuristics. The scenario where no clustering heuristics are applied is referred to as \emph{None}, with each address being considered as a separate entity. The most significant decrease is seen with the result of the CS heuristic, reducing 1,011,406 entities, or about 2.3\% of the total number of entities. The least significant decrease is seen with the result of the CS+G heuristic, reducing 568,422 entities, or about 1.6\% of the total number of entities. This highlights that unconfirmed transactions contain additional information abouy inter-address association that can enhance address clustering.

We summarize that the clustering result of unconfirmed transactions improves the clustering result of confirmed transactions from four perspectives. The first improvement \emph{ADD} refers to the clustering result of unconfirmed transactions expanding the clustering result of confirmed transactions, i.e., they expand the number of addresses in a given cluster. The second improvement \emph{MERGE} refers to the clustering result of unconfirmed transactions merging multiple clusters of confirmed transactions into one cluster. The third improvement \emph{SUB} refers to a cluster of unconfirmed transactions being the subset of a cluster of confirmed transactions, which increases the credibility of the clustering result of confirmed transactions. The fourth improvement \emph{NEW} refers to none of the addresses in the clusters of unconfirmed transactions appearing in the clusters of confirmed transactions.

\begin{table}[t]
\centering
\caption{Improvement of the SU clustering result on the SC clustering result.}
\label{table:enhancement}
\scalebox{0.8}{
\begin{tabular}{lrrrrr} 
\toprule
\textbf{Heuristic}            & \multicolumn{1}{l}{\textbf{Clusters}} & \multicolumn{1}{l}{\textbf{ADD}} & \multicolumn{1}{l}{\textbf{MERGE}} & \multicolumn{1}{l}{\textbf{SUBSET}} &  \multicolumn{1}{l}{\textbf{NEW}}  \\ 
\hline
CS                   & 416,172                               & 348                              & 3,761                              & 407,950                                                                & 4,113                            \\
CS+A  & 393,792                               & 43,289                           & 30,681                             & 316,367                                                              & 3,455                            \\
CS+M & 382,103                               & 134,946                          & 23,618                             & 221,742                                                               & 1,797                            \\
CS+G  & 382,092                               & 134,321                          & 26,845                             & 219,049                                                               & 1,797                            \\
CS+E    & 396,726                               & 35,813                            & 27,758                             & 329,479                                                              & 3,676                            \\
CS+K    & 388,537                               & 33,455                            & 32,913                             & 319,628                                                              & 2,541                            \\
\bottomrule
\end{tabular}
}
\label{table:existing on unconfirmed}
\end{table}

Table~\ref{table:existing on unconfirmed} shows the impact of applying the state-of-the-art clustering heuristics to unconfirmed transactions. In the case that the state-of-the-art clustering heuristic is considered accurate, \emph{MERGE} clusters can merge multiple SC clusters belonging to the same entity. \emph{SUBSET} clusters can enhance the credibility of the clustering result of the confirmed transactions. Although there are addresses in \emph{ADD} and \emph{NEW} clusters that haven't been involved in confirmed transactions, users may still use these addresses to participate in transactions in the future. Therefore, clusters that include addresses that have not been involved in confirmed transactions are also important for analyzing user behavior and revealing their privacy.

These clustering heuristics are not limited to merely clustering Bitcoin addresses controlled by a single entity, but also to monitor the movement of bitcoins as they are transferred between different entities. Unconfirmed transactions in the mempool can improve the clustering result of confirmed transactions, leading to a more accurate fund flow tracking. Table~\ref{table:enhancement} demonstrates that the clustering result of unconfirmed transactions can identify more addresses belonging to a certain entity, or reduce the error that addresses belonging to the same entity are divided into two clusters. In this case, the analysis of the user behavior will be more accurate.

\section{Novel Clustering Heuristics}
\label{sec:novel}


In this section, we concentrate on unconfirmed transactions, particularly failed transactions, which have received limited attention in previous research. We propose three novel clustering heuristics by examining specific user behavior patterns embodied in unconfirmed transactions. Subsequently, we demonstrate the effect of these novel heuristics on the clustering result.

\subsection{Definition of Novel Clustering Heuristics}

During the exploration, we discover two interesting phenomena in unconfirmed transactions. The first phenomenon is that users replace old unconfirmed transactions by initiating a new transaction with a higher fee. This mechanism, called Replace-By-Fee (RBF), allows the sender to replace a transaction with another transaction that pays a higher fee if the network becomes congested. If a user finds that a transaction s/he initiated has not been confirmed for a long time, or has initiated an incorrect transaction, s/he can choose to initiate a new transaction with a higher fee to replace it. Additionally, when a transaction is initiated, the user must specify whether the transaction can be replaced by setting the \emph{replaceable} field mentioned by Table~\ref{table:extra_details} in the transaction.

The second interesting phenomenon is first discovered in failed transactions. Bitcoin mempool is designed to accept transactions spending UTXOs of unconfirmed transactions, and miners may package these transactions into a block at the same time. We refer to the dependency chain as a sequence of unconfirmed transactions in which one unconfirmed transaction utilizes the UTXO of another unconfirmed transaction, which in turn spends the UTXO of another unconfirmed transaction, and so on. The user must spend the UTXO of confirmed transactions, so both parties of the transaction in the dependency chain either negotiate in advance so that the receiver believes that s/he will receive a spendable UTXO for initiating the next transaction, or both parties to the transaction are actually the same user. If the first transaction in the dependency chain is not confirmed, then all subsequent transactions will fail. Confirmed transactions may also form such a dependency chain while they are in the mempool, but it is not possible to reproduce this dependency chain through information in the blockchain.

\begin{figure}[htbp]
    \centering
    \includegraphics[scale=0.35]{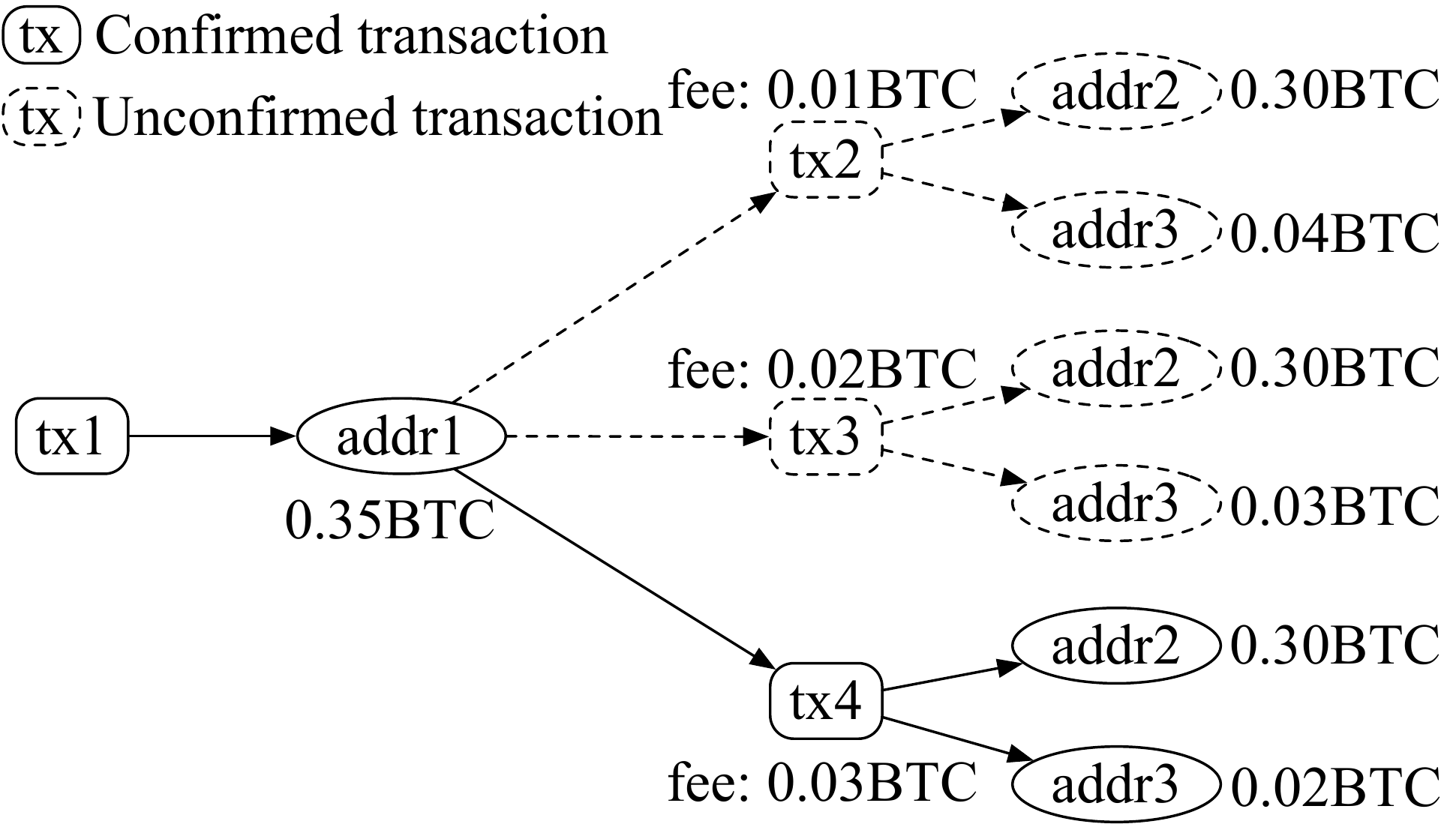}
    \caption{Illustration of heuristic \emph{replacement change}.}
    \label{fig:heuristic2}
\end{figure}

\begin{figure}[htbp]
    \centering
    \includegraphics[scale=0.38]{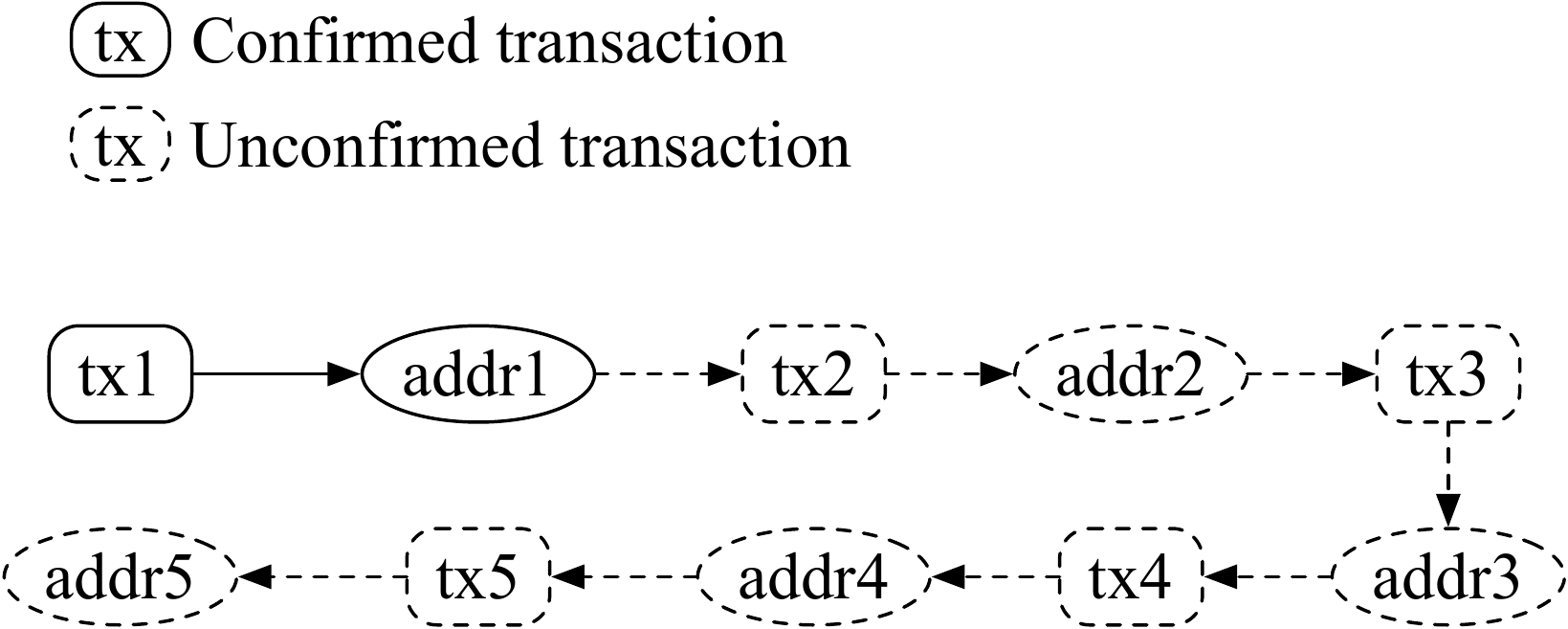}
    \caption{Illustration of heuristic \emph{one-to-one chain}.}
    \label{fig:heuristic3}
\end{figure}

For the first phenomenon, we designed one clustering heuristic to determine the change address.

\smallskip\noindent
\textbf{Heuristic \emph{replacement change}.} In multiple replaceable transactions that attempt to spend the same UTXO, the output address is a change address if its received amount changes in different transactions.

In an actual transaction, the amount of the transaction is agreed upon between the parties and does not change casually. When the user raises the transaction fee, the amount paid to the other party remains the same, while the amount received by the change address decreases. As shown in Figure~\ref{fig:heuristic2}, the transaction input amounts for tx2, tx3, and tx4 are all 0.35 BTC. When the transaction fee increases, the amount received by addr2 remains the same, and the amount received by addr3 decreases. Therefore, we consider addr2, which receives the same amount all the time, as the counterparty address, and addr3, which receives an ever-smaller amount, as the sender's change address.

For the second phenomenon, we designed two clustering heuristics against two special transaction patterns respectively in the dependency chain.

\smallskip\noindent
\textbf{Heuristic \emph{one-to-one chain}.} Addresses in a one-to-one chain containing more than 2 transactions are controlled by the same entity.

\begin{figure}[t]
    \centering
    \includegraphics[width=7cm]{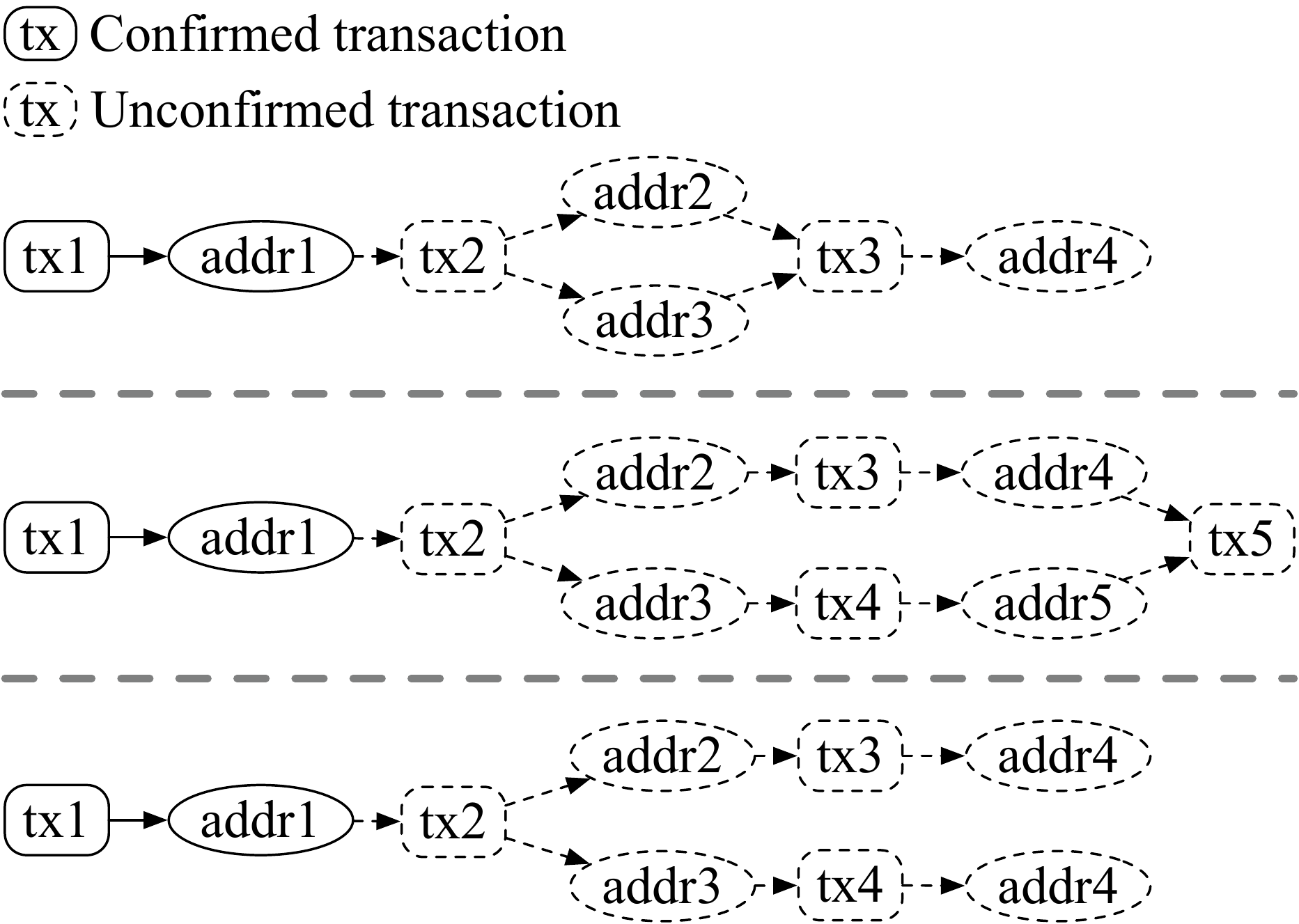}
    \caption{Illustration of heuristic \emph{fusiform chain}.}
    \label{fig:heuristic5}
\end{figure}

The definition of the one-to-one chain is that every transaction in the chain has only one input and one output, and each transaction in the one-to-one chain spends the UTXO of another unconfirmed transaction. To avoid false positives caused by prior agreements between the two parties, we have imposed a stricter limit on this heuristic, requiring containing more than 2 transactions. There are two main reasons for this. One is that users want miners to package multiple transactions of theirs at once, without having to wait for each transaction to be confirmed before initiating another one. The initiation times of transactions in a one-to-one chain are very close, usually under ten minutes. The other reason is that a user initiates a high-fee transaction spending the UTXO(s) of a long-time unconfirmed transaction, hoping that miners will package the first transaction for the sake of the high fee of the second transaction. This transaction model is known as Child-Pays-for-Parent~(CPFP), where the child transaction compensates the parent~(ancestor) transaction so that both can be confirmed more quickly. In this scenario, an unconfirmed transaction with a high fee can incentivize miners to mine its ancestor transactions that are also unconfirmed, resulting in multiple unconfirmed transactions being confirmed at the same time. These transactions are initiated at a relatively large interval, usually greater than ten minutes.

As shown in Figure~\ref{fig:heuristic3}, tx2, tx3, tx4, and tx5 are all unconfirmed transactions whose initiation times are very close. All of tx3, tx4, and tx5 spend UTXOs of unconfirmed transactions, reflecting the user's intention to have multiple transactions confirmed within the time of a block. As a result, we can consider that addr1, addr2, addr3, addr4, and addr5 are controlled by the same entity.

\smallskip\noindent
\textbf{Heuristic \emph{fusiform chain}.} Addresses in the fusiform chain are controlled by the same entity.



We define the fusiform chain as a transaction pattern in which multiple UTXOs generated by a previous transaction are aggregated to the same transaction or address. Alarab et al.~\cite{alarab2020comparative} use supervised learning methods to detect money laundering transactions in Bitcoin and find that fusiform transactions are a common pattern in such transactions. Möser et al.~\cite{DBLP:conf/ecrime/MoserBB13} conduct Bitcoin-related anti-money laundering research. Through the study of the transaction strategies of service providers who provide anonymous transactions, this work finds that the fusiform structure is the main method used by such services to enhance the anonymity of transactions. The primary reason users initiate fusiform transactions is to enhance the anonymity of the transaction.


\begin{figure}[htbp]
    \centering
    \includegraphics[scale=0.55]{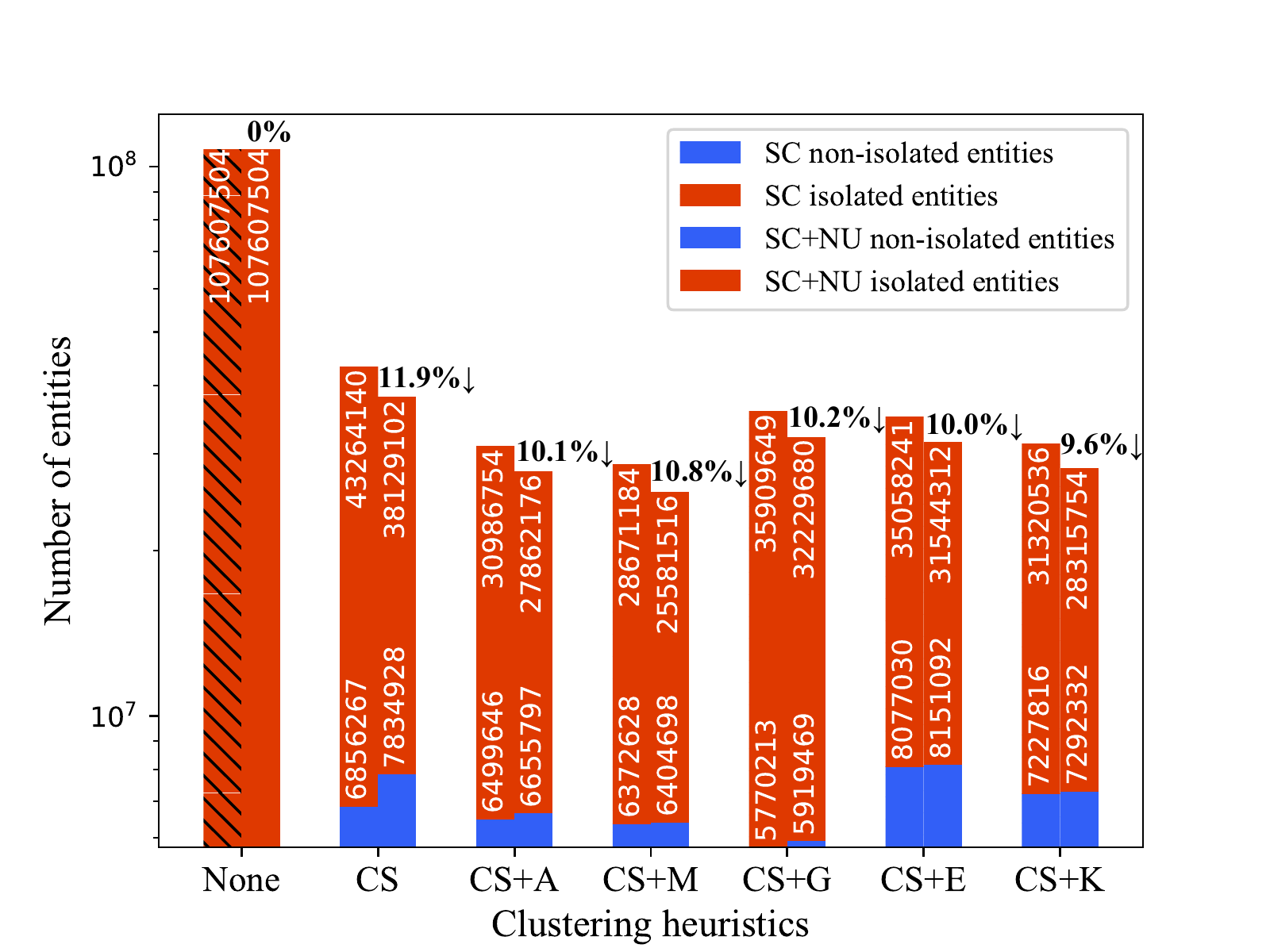}
    \caption{Number of entities in the SC clustering result and the SC+NU clustering result.}
    \label{fig:compare_2}
\end{figure}

Figure~\ref{fig:heuristic5} shows three typical fusiform chains. In the first case, tx2 generates two UTXOs that are then combined in tx3. This is consistent with the result of the co-spend heuristic, where addr2 and addr3 are controlled by the same entity. In the second case, there are two transactions~(tx3 and tx4) between tx2 and tx5, but the UTXOs generated by tx2 are finally combined to tx5. Therefore, this heuristic considers that addr2, addr3, addr4, and addr5 are controlled by the same entity. In the third case, two UTXOs generated by tx2 end up transferring bitcoins to addr4 through tx3 and tx4, respectively. We also take this case as a fusiform chain, i.e. the fusiform chain ends at the same address. Therefore, addr2, addr3, and addr4 are controlled by the same entity.

\begin{table}[t]
\centering
\caption{Improvement of the NU clustering result on the SC clustering result.}
\scalebox{0.8}{
\begin{tabular}{lrrrrrr} 
\toprule
\textbf{Heuristic}            & \multicolumn{1}{l}{\textbf{NU Clusters}} & \multicolumn{1}{l}{\textbf{ADD}} & \multicolumn{1}{l}{\textbf{MERGE}} & \multicolumn{1}{l}{\textbf{SUBSET}} &  \multicolumn{1}{l}{\textbf{NEW}}  \\ 
\hline
CS                   & 2,531,354                               & 2,933                              & 2,034,536                              & 492,687                                                                 & 1,198                            \\
CS+A & 2,531,354                               & 3,027                           & 1,541,223                             & 985,907                                                              & 1,197                            \\
CS+M & 2,531,354                               & 3,300                          & 1,540,380                             & 986,479                                                               & 1,195                            \\
CS+G  & 2,531,354                              & 3,255                          & 1,575,390                             & 951,515                                                               & 1,194                            \\
CS+E    & 2,531,354                              & 2,968                            & 1,628,029                             & 899,159                                                              & 1,198                            \\
CS+K    & 2,531,354                              & 2,817                            & 1,603,867                             & 923,471                                                              & 1,199                            \\

\bottomrule
\end{tabular}
}
\label{table:enhancement-offchain-not-merge}
\end{table}

Although the dependency chain formed by confirmed transactions while they are in the mempool cannot be reproduced from information in the blockchain alone, we can record the state of confirmed transactions when in the mempool and reproduce the dependency chain through the \emph{depend} and \emph{spentby} fields in Table~\ref{table:extra_details}. As a result, the latter two novel heuristics can also be applied to this scenario.

\subsection{Evaluation of Novel Clustering Heuristics}

To show the influence of the above clustering heuristics, we apply them to unconfirmed transactions. For the convenience of description, we refer to the result of applying three novel clustering heuristics to unconfirmed transactions as the NU clustering result. Then, we analyze the impact of the NU clustering result on the SC and SC+SU clustering results.

\begin{figure}[htbp]
    \centering 
    \includegraphics[scale=0.55]{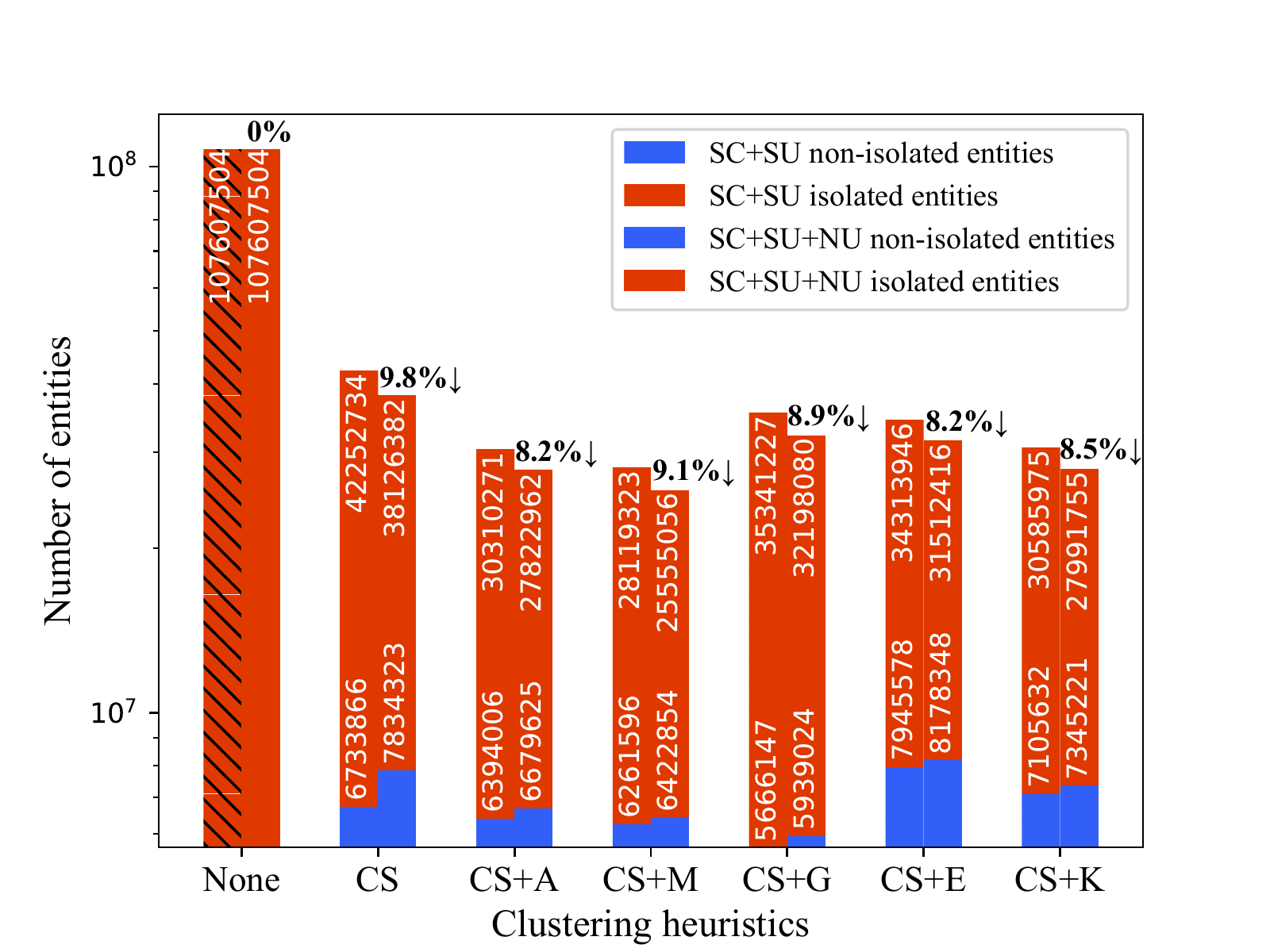}
    \caption{Number of entities in the SC+SU clustering result and the SC+SU+NU clustering result.}
    \label{fig:compare_3}
\end{figure}


Figure~\ref{fig:compare_2} demonstrates that the novel heuristics are able to mine the association information between addresses contained in unconfirmed transactions, resulting in the identification of more addresses that belong to the same entity. More addresses or clusters belonging to the same entity are clustered together. Therefore, the number of entities in the SC+NU clustering result is less than the number of entities in the origin clustering result.


Figure~\ref{fig:compare_3} demonstrates that unconfirmed transactions contain additional inter-address association information, which can improve the clustering results. We design three new clustering heuristics specifically for unconfirmed transactions. This is because user behavior in unconfirmed transactions differs from that in confirmed transactions, making it challenging to discover associations using state-of-the-art clustering heuristics. The results demonstrate that our novel clustering heuristics are more effective in mining associations between addresses compared to the state-of-the-art clustering heuristics. This leads to a further decrease in the number of entities in the SC+SU+NU clustering result, with the co-spend clustering result being the most affected, reducing the number of entities by 4,126,352 (9.8\% of the total). The clustering result of the CS+M clustering heuristic is least affected, with a reduction of 2,801,530 entities (8.2\% of the total).

Table~\ref{table:enhancement-offchain-not-merge} shows that the most frequent category in each clustering result is \emph{SUBSET}, indicating that the clustering result of the novel clustering heuristics is consistent with that generated by the state-of-the-art clustering heuristics. This implies that the clustering result generated by novel clustering heuristics is accurate.


\begin{table}[t]
\centering
\caption{Improvement of the NU clustering result on the SC+SU clustering result.}
\scalebox{0.8}{
\begin{tabular}{lrrrrrr} 
\toprule
\textbf{Heuristic}            & \multicolumn{1}{l}{\textbf{NU Clusters}} & \multicolumn{1}{l}{\textbf{ADD}} & \multicolumn{1}{l}{\textbf{MERGE}} & \multicolumn{1}{l}{\textbf{SUBSET}} & \multicolumn{1}{l}{\textbf{NEW}}  \\ 
\hline
CS                   & 2,531,354                               & 931                              & 2,034,432                              & 495,877                                                                 & 114                            \\
CS+A & 2,531,354                               & 563                           & 1,538,947                             & 992,634                                                              & 110                            \\
CS+M & 2,531,354                               & 573                          & 1,536,497                             & 994,266                                                               & 18                            \\
CS+G  & 2,531,354                              & 571                          & 1,568,750                             & 962,015                                                               & 18                            \\
CS+E    & 2,531,354                              & 644                            & 1,538,047                             & 906,062                                                              & 110                            \\
CS+K    & 2,531,354                              & 595                            & 1,537,532                             & 993,118                                                              & 109                            \\
\bottomrule
\end{tabular}
}
\label{table:enhancement-offchain}
\end{table}

\begin{figure*}[htbp]
\centering
\subfigure[Heuristic replacement change]{
\begin{minipage}[t]{4.8cm}
\centering
\includegraphics[width=4.8cm]{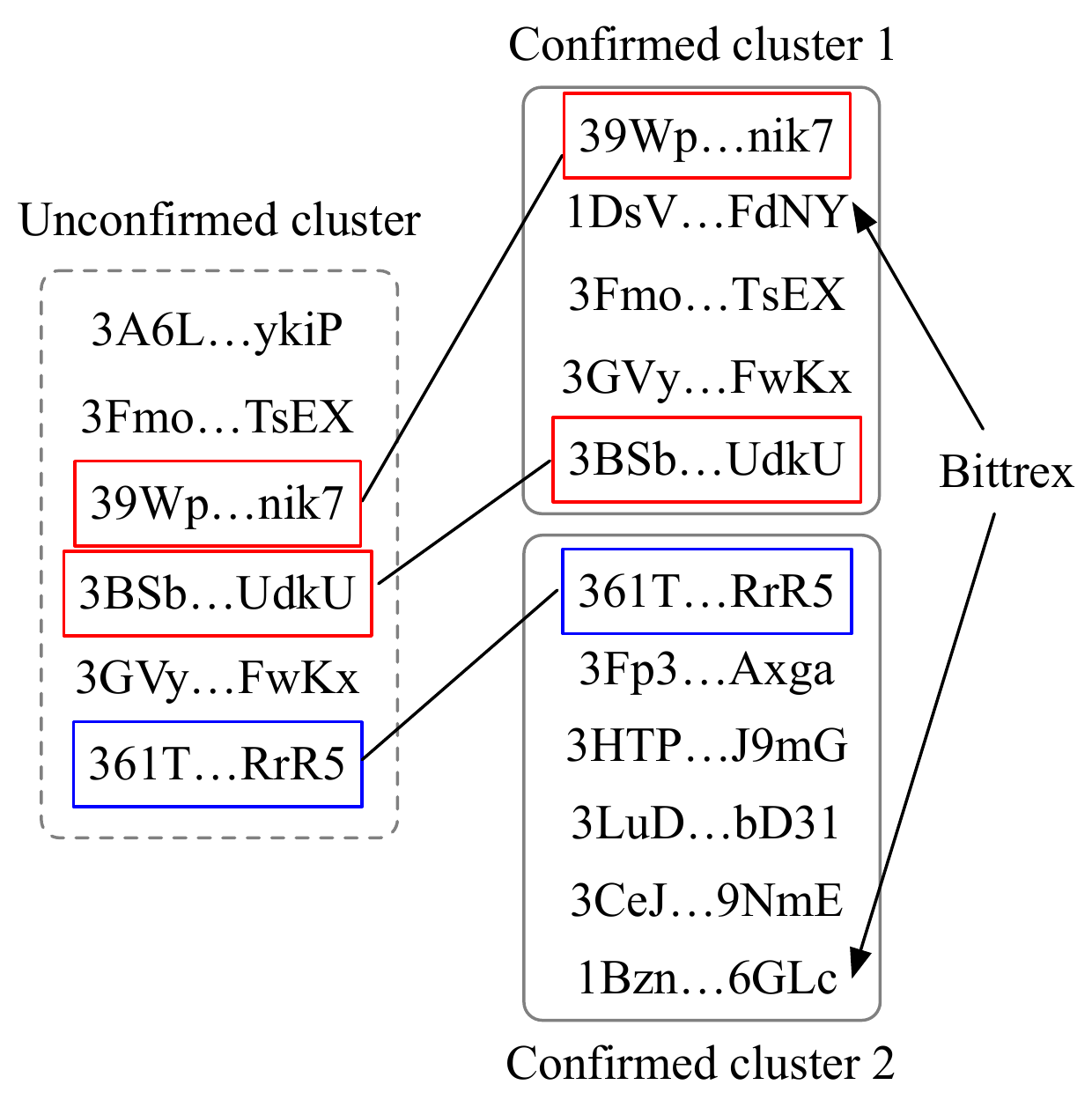}
\label{fig:case1}
\end{minipage}
}%
\hspace{1mm}
\subfigure[Heuristic one-to-one chain]{
\begin{minipage}[t]{5.7cm}
\centering
\includegraphics[width=5.7cm]{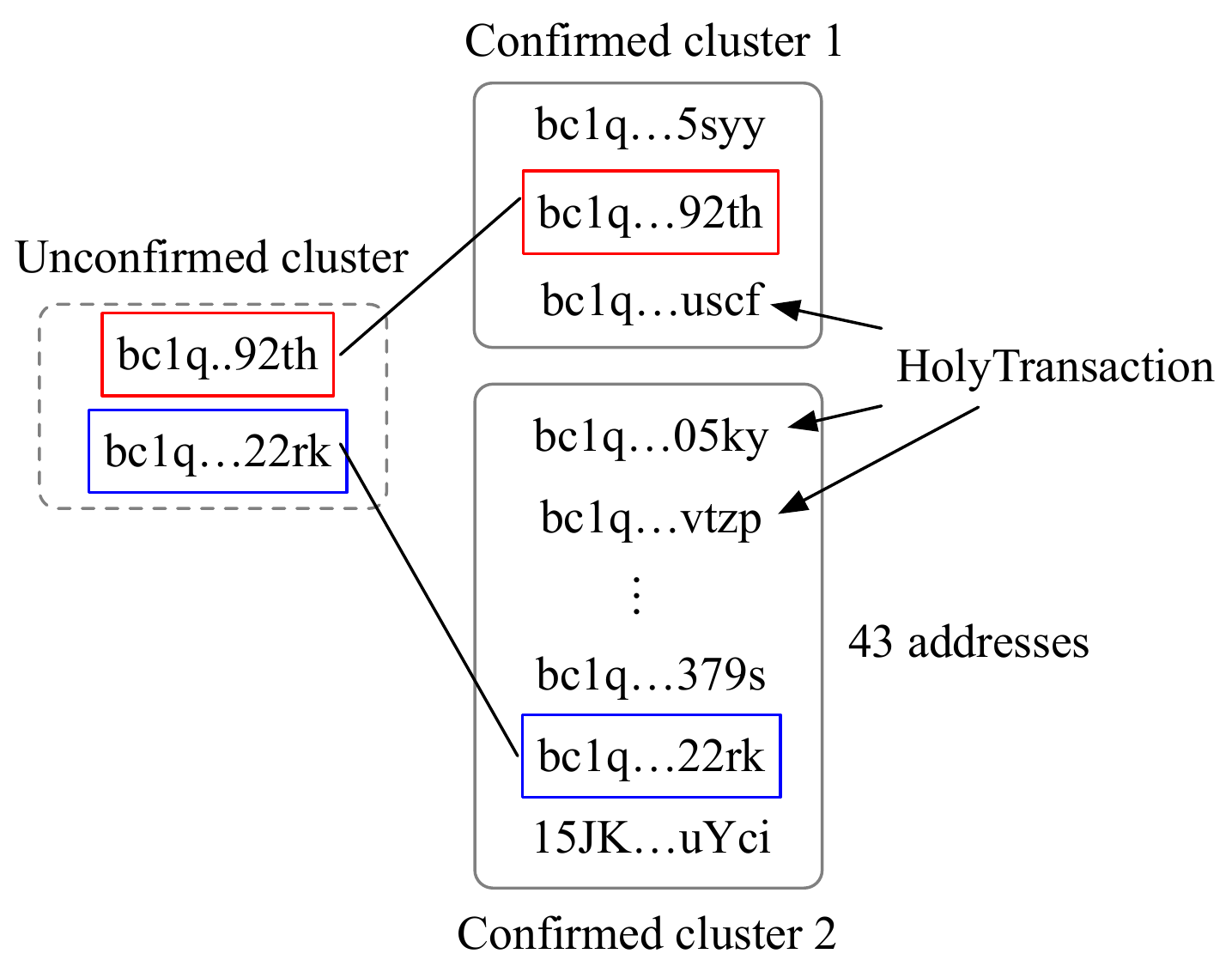}
\label{fig:case2}
\end{minipage}
}%
\hspace{1mm}
\subfigure[Heuristic fusiform chain]{
\begin{minipage}[t]{5.7cm}
\centering
\includegraphics[width=5.7cm]{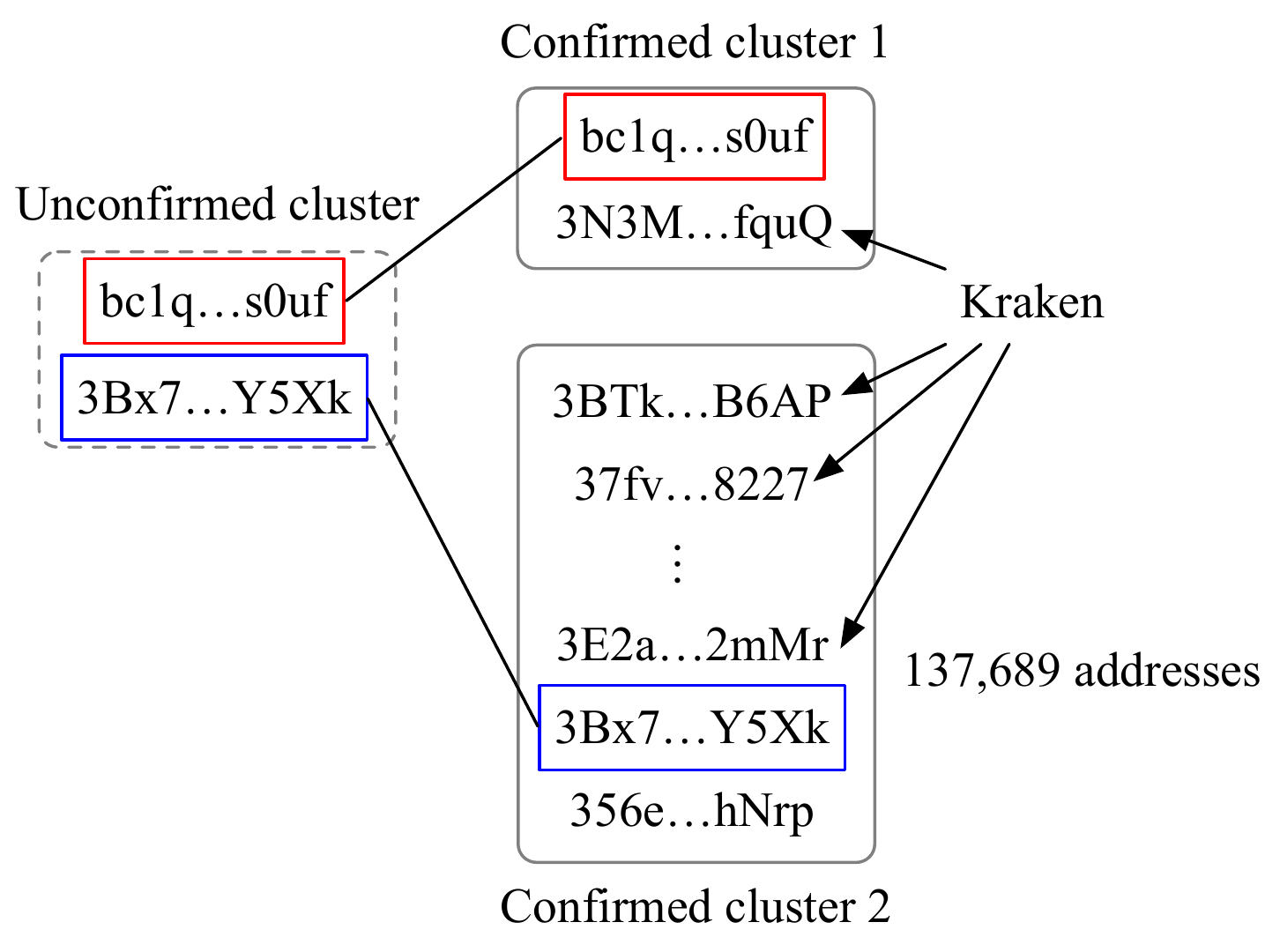}
\label{fig:case3}
\end{minipage}%
}%
\vspace{-3mm}
\caption{Three cases of the unconfirmed cluster merging two confirmed clusters with the same label. Unconfirmed clusters are generated by applying novel clustering heuristics to unconfirmed transactions. Confirmed clusters are generated by applying the state-of-the-art clustering heuristics to confirmed transactions.}
\label{fig:cases}
\end{figure*}

The number of \emph{SUBSET}s in Table~\ref{table:enhancement-offchain} is larger than those in Table~\ref{table:enhancement-offchain-not-merge}, which indicates that the clustering result of the novel clustering heuristics on unconfirmed transactions is partially consistent to the results of the state-of-the-art clustering heuristics on unconfirmed transactions. A large number of \emph{MERGE}s illustrate that the novel clustering heuristics can exploit unconfirmed transactions and are better at mining the associations between addresses from unconfirmed transactions compared to the state-of-the-art clustering heuristics. The presence of \emph{ADD} and \emph{NEW} also supports this observation, although there are fewer of these instances.

To validate the novel clustering heuristics, we collect address labeled data from the WalletExplorer website~\cite{walletexplorer}. With the limited labeled data, we find that the clustering result of the novel clustering heuristics on unconfirmed transactions is able to merge multiple clusters of the same entity. That is, our novel clustering heuristics correct the mistake of the state-of-the-art clustering heuristics, which wrongly cluster addresses that belong to the same user into two separate clusters. Figure~\ref{fig:cases} presents a selected case for each novel clustering heuristic. For instance, Figure~\ref{fig:case1} shows that a cluster of 6 addresses generated by the Heuristic \emph{replacement change} merges two clusters belonging to the Bittrex exchange. These cases provide evidence of the effectiveness of the novel clustering heuristics with limited labeled data.

\vspace{-2mm}
\section{Discussion}
\label{sec:discussion}
\vspace{-1mm}
\subsection{Insufficient Validation Data} 

Since our work is the first to systematically study Bitcoin mempool to improve address clustering, we are missing a ground truth against which we can compare our results.

High-quality labeled data is critical for Bitcoin data analytics but it is difficult to obtain due to the anonymity protection provided by Bitcoin. Although some blockchain data analytics companies may have relevant data, the information they rely on is not publicly disclosed.

We collect Bitcoin addresses of known entities from the WalletExplorer.com website~\cite{walletexplorer} that is widely used as the ground truth in several studies~\cite{foley2019sex, DBLP:journals/cybersecurity/Paquet-Clouston19,DBLP:journals/corr/abs-1906-07852,DBLP:journals/apin/ZolaSBGU22,DBLP:conf/iscisc/ManaviH20}. The site divides entities into five categories: exchanges, mining pools, service providers, gambling, and darknet marketplaces. This site claims it follows a basic algorithm to determine entity addresses. Addresses are merged together, if they are co-spend in one transaction, i.e., the co-spend heuristic. So if addresses A and B are co-spend in the transaction tx1, and addresses B and C are co-spend in the transaction tx2, all addresses A, B, and C will be part of the same entity. However, the name database of this website is not updated~(except some very rare cases) since 2016. The labeled addresses are now expanded by the co-spend heuristic and do not take into account the effect of mixing transactions. Therefore, this labeled address dataset is not guaranteed to be completely correct. And we start collecting mempool data on May 1, 2022, so the labeled addresses on the website rarely appear in our collection.


\begin{figure}[t]
    \centering
    \includegraphics[scale=0.36]{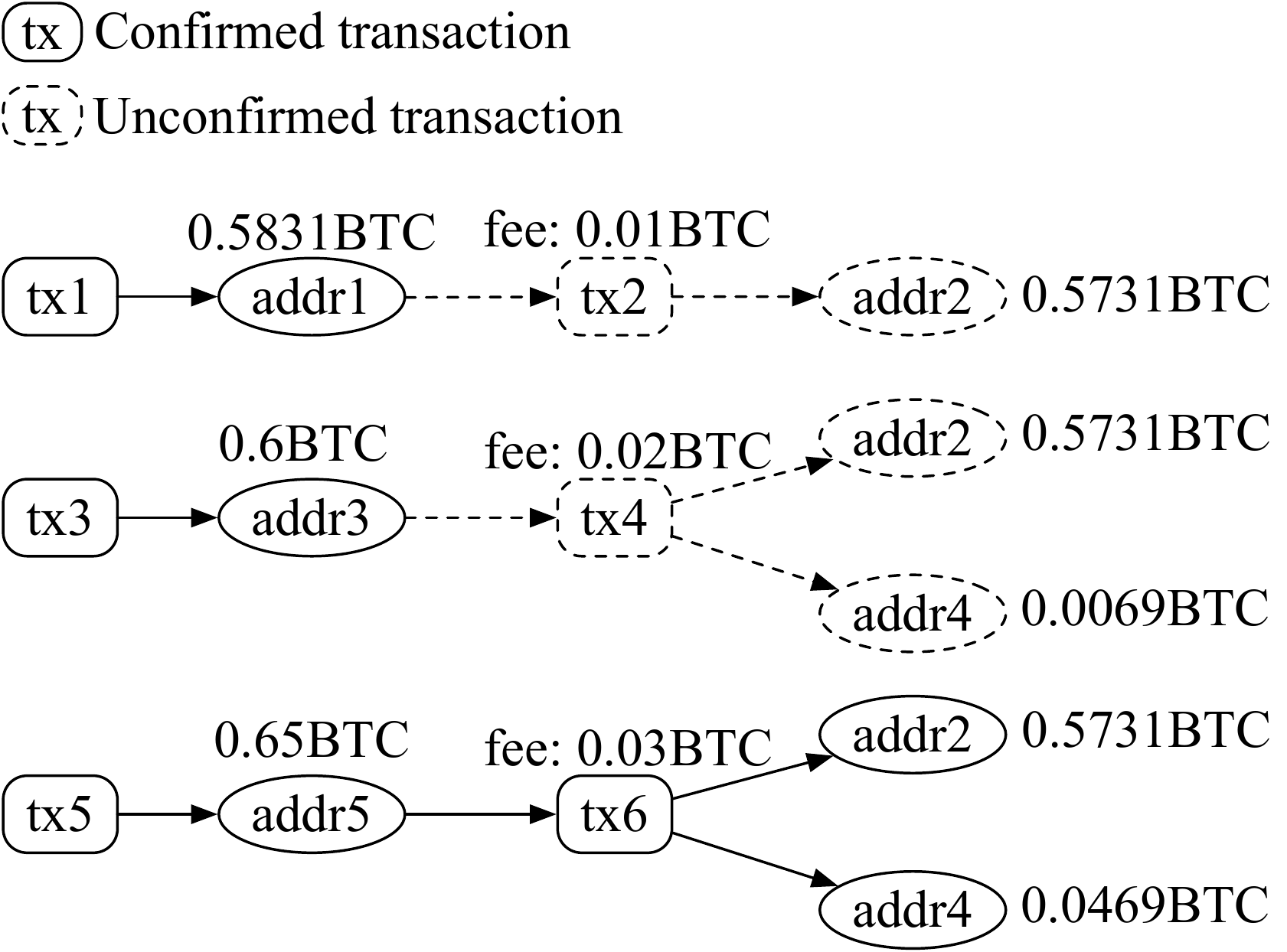}
    \caption{Illustration of heuristic \emph{new input}.}
    \label{fig:newinput}
\end{figure}

Previous studies on address clustering utilize historical Bitcoin transactions as a dataset, which has a large number of labeled data. Metrics such as accuracy and recall are calculated using these labeled data to evaluate clustering methods. However, as the transaction data used in our work is relatively new, there is not sufficient labeled data to support us in carrying out the statistical validation. However, the novel clustering heuristics proposed in Section~\ref{sec:novel} consider several particular transaction behaviors as constraints, leading us to believe that the results obtained of these novel heuristics are accurate. This is further supported by our validation of the clustering results with three labeled cases in Section~\ref{sec:novel}. In the next section, we present two additional clustering heuristics that require further validation using more labeled data.

\subsection{Additional Clustering Heuristics}
In this section, we present two additional clustering heuristics that require validation with more labeled data. The first one aims to enhance the peel chain heuristic, while the second one focuses on investigating replacement transactions. These two clustering heuristics may produce false positives due to the presence of special cases. Therefore, these two clustering heuristics require some additional parameters to reduce the false positive rate, which requires a large amount of labeled data for analysis. The labeled data we have collected so far is not enough to support us in validating and refining these two clustering heuristics.

\smallskip\noindent
\textbf{Heuristic \emph{new input}.} In multiple unconfirmed transactions where the same address serves as the recipient, if the amount received by this address is exactly identical, it can be inferred that transaction inputs for these multiple transactions are controlled by the same entity.


When the user increases the transaction fee, the wallet software may select the new UTXO~(s) with a larger balance to replace the original UTXO~(s) if the balance of the original UTXO~(s) is insufficient to meet the transaction format requirements. Therefore, the new UTXO~(s) and the original UTXO~(s) are controlled by the same entity. As shown in Figure~\ref{fig:newinput}, in transactions tx2, tx4, and tx6, the sender transfers an identical amount of bitcoins to addr2. Initially, the user uses addr1 to make a payment to addr2 in  tx2 that is not confirmed for a long time. In the process of increasing the transaction fee, the user or wallet software selects addr3 to pay to addr2 in tx4 and selects addr5 to pay to addr2 in tx6. Moreover, addr2 receives exactly the same amount in each transaction. Therefore, this heuristic considers that addr1, addr3, and addr5 are controlled by the same entity. However, there may be special cases where this heuristic fails. For example, addr2 is a merchant address. addr1, addr3, and addr5 are controlled by three separate entities making a purchase from addr2 but only the transaction tx6 is confirmed. Therefore, this heuristic needs more labeled data to be validated.

\smallskip\noindent
\textbf{Heuristic \emph{peel chain}.} Transactions in a peel chain are all initiated by the same entity, i.e., inputs of transactions in a peel chain are all controlled by the same entity.

A peel chain is a transaction pattern in the blockchain. Meiklejohn et al.~\cite{DBLP:conf/imc/MeiklejohnPJLMVS13} introduced the concept of a peel chain, which is a sequence of transactions that originates from a transaction with a UTXO associated with a large amount of bitcoins as input. The spending of this UTXO produces two outputs: one that represents a payment to a different entity, which is relatively small, and another that indicates the change, which is comparatively large. This procedure can be iterated several times where each step in the chain gradually peels off smaller values until the remaining change amount becomes small.~\cite{DBLP:conf/imc/MeiklejohnPJLMVS13}. In Bitcoin mempool, there are peel chains containing unconfirmed transactions. Kappos~\cite{DBLP:conf/uss/KapposYSRHM22} extends the concept of peel chains to include more general scenarios, where transactions may have multiple inputs and outputs. Furthermore, they require that every consecutive transaction in the peel chain must be connected by a change output. Through our observation, peel chains consisting of unconfirmed transactions always have one input and two outputs.

\begin{figure}[t]
    \centering
    \includegraphics[scale=0.35]{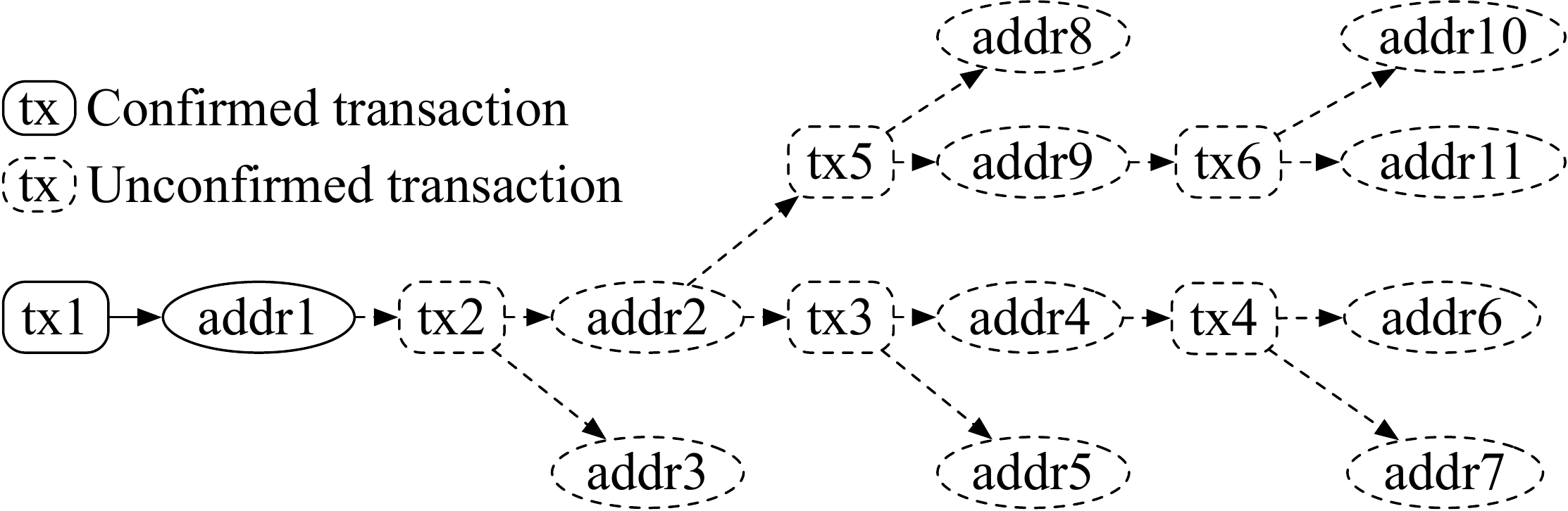}
    \caption{Illustration of heuristic \emph{peel chain}.}
    \label{fig:heuristic4}
\end{figure}

Figure~\ref{fig:heuristic4} shows two peel chains in the mempool. First, tx1, tx2, and tx3 make up a peel chain where every transaction has one input and two outputs. All of tx1, tx2, and tx3 are initiated by the same entity who repeats the peel chain pattern three times to peel small value from addr1. Therefore, addr2, addr4, addr6 are controlled by the same entity. Second, tx2, tx5, and tx6 consist of another peel chain where addr2, addr9, addr11 are controlled by the same entity.




\vspace{-1mm}
\subsection{Mempool Optimization for Privacy} The analysis in Section~\ref{sec:analysis} demonstrates that data in the mempool can leak user privacy, which is contrary to the purpose of Bitcoin to protect user privacy. Therefore, improvements are needed for Bitcoin mempool so that transaction details, such as transaction inputs and outputs, transaction amounts, etc., cannot be observed until transactions are confirmed.


First, we can refer to the zero-knowledge proofs in Zcash~\cite{DBLP:conf/uss/KapposYMM18} to enhance privacy through hiding transaction details. It achieves not only the hiding of the sender and receiver addresses but also the hiding of the transaction amount, only the two parties of the transaction can be linked to the transaction and the amount of the transaction, while the data viewed by other users can only verify the validity of the transaction and the correctness of the amount, but cannot get other information such as the two sides of the transaction and the amount of the transaction, achieving a high level of strong privacy protection. This approach makes the transaction details visible only to the two parties of the transaction in both the mempool and the blockchain, but also undermines the principle of openness of the ledger.

Second, we can modify the Bitcoin transaction structure by adding a new field that flags whether the transaction can be forwarded by other Bitcoin nodes. If a user wants to hide his/her transaction behavior in the mempool, s/he can construct a transaction with this field and only send this transaction to miner nodes s/he trusts. In this case, the miner node will only put this transaction into its own mempool waiting to be packed, and will not forward it to other Bitcoin nodes. This makes a transaction invisible in the mempool but visible in the blockchain so failed transactions will not be accessible to other nodes except miner nodes, thus protecting user privacy. However, this approach requires that miners do not disclose these unconfirmed transactions. In addition, as the number of miners the user sends transactions to increases, the transaction confirmation may be faster, but the privacy of users decreases.


\section{Related Work}
\label{sec:relateworks}

Our study is closely related to the literature on Bitcoin address clustering and Bitcoin mempool analysis.

\subsection{Bitcoin Address Clustering}
The anonymity of Bitcoin makes it difficult to determine the ownership of Bitcoin addresses. Several methods are proposed to cluster associated Bitcoin addresses by utilizing heuristics.

\smallskip
\noindent\emph{Co-spend methods}. At first, researchers focus on the relationship between the transaction inputs. The co-spend method~\cite{DBLP:conf/imc/MeiklejohnPJLMVS13} considers all inputs of a transaction are controlled by the same entity. Cazabet et al.~\cite{DBLP:conf/complexnetworks/RemyRM17} point out that the original co-spend method has a relatively low recall for users. On the basis of the co-spend heuristic, Kalodner et al.~\cite{DBLP:journals/corr/abs-1709-02489} propose to reduce clustering interference caused by a special type of mixing transactions, i.e., ${\it CoinJoin}$. The Coinjoin detection method in Blocksci~\cite{DBLP:journals/corr/abs-1709-02489} has been widely applied in address clustering. However, the improved method only solves excessive clustering introduced by a certain type of mixing transactions but not mines the potential association of Bitcoin addresses from these transactions, which somehow reduces the recall of clustering results.

\smallskip\noindent
\emph{Change address methods}. In addition to studying the associations between transaction inputs, a few methods~\cite{DBLP:conf/fc/AndroulakiKRSC13,DBLP:conf/imc/MeiklejohnPJLMVS13, DBLP:journals/popets/GoldfederKRN18, DBLP:conf/icmla/ErmilovPY17, DBLP:conf/uss/KapposYSRHM22,DBLP:journals/access/Zhang0020} with different patterns are proposed to study the association of output addresses. First, researchers~\cite{DBLP:conf/fc/AndroulakiKRSC13,DBLP:conf/imc/MeiklejohnPJLMVS13, DBLP:journals/popets/GoldfederKRN18} mainly consider the behavior of the transaction outputs. They all require the change address is a fresh address, meaning that this is the first time the address appears in the blockchain. Then, Ermilov et al.~\cite{DBLP:conf/icmla/ErmilovPY17} consider not only the behavior of the transaction outputs but also the value they received. It requires the value of the change address is significant to at least the fourth decimal place. These patterns have been extended to analyze the anonymity of other cryptocurrencies and help cluster their associated addresses, such as Zcash~\cite{DBLP:conf/uss/KapposYMM18} and Ripple~\cite{DBLP:journals/popets/Moreno-SanchezZ16}. Recently, Kappos et al.~\cite{DBLP:conf/uss/KapposYSRHM22} don't just focus on one transaction. It considers a specific transaction type called \emph{peel chain} that contains a series of transactions.

However, some transactions in Bitcoin may mismatch these patterns and result in incorrect Bitcoin associations. For example, the fresh address rule considers the new address in the outputs of a two-output transaction as the change address of the sender. The new output address in the ransom payment transactions of  {\em Locky} is actually the unique ransom address generated by the criminals for every victim, rather than a change address of a victim~\cite{DBLP:conf/sp/HuangALIBMLLSM18,DBLP:conf/raid/MehnazMB18,DBLP:conf/iconip/McIntoshJWS19}.

Co-spend methods and change address methods focus on confirmed transactions in the blockchain. To enlarge the cluster generated by confirmed transactions, we design a series of novel clustering heuristics by combining confirmed transactions and failed transactions.

\vspace{-2mm}
\subsection{Bitcoin Mempool Analysis}

Recently, researchers have also focused on mempools, but related studies are still relatively few. They focus on two main areas: prediction of the transaction confirmation time and analysis of attacks against mempools.

\smallskip\noindent
\emph{Prediction of the transaction confirmation time}. As Bitcoin is universally recognized as the most popular cryptocurrency, more and more Bitcoin transactions are expected to be populated to the Bitcoin blockchain system. However, transactions cannot be confirmed altogether into the next block due to the limited block capacity. One of the most demanding requirements for users to use Bitcoin is to estimate the confirmation time of a newly submitted transaction.

Several works~\cite{DBLP:conf/blocksys/KoJMLH19, DBLP:journals/sigmetrics/StoepkerGK21, DBLP:journals/sigmetrics/GundlachGKR21,
DBLP:journals/fgcs/PierroRDMT22,
DBLP:conf/wise/ZhangZLXL22} propose methods to estimate the confirmation time for a single transaction. Gundlach et al.~\cite{DBLP:journals/sigmetrics/GundlachGKR21} study the distribution of confirmation times of Bitcoin transactions, conditional on the size of the current mempool. They propose a unique method based on rigorous mathematical analysis and predict the confirmation times of Bitcoin transactions through a Cramer-Lundberg~(CL) model. Their model shows how the transaction confirmation times, when modeled through the CL model and inverse Gaussian distribution does not significantly change after c=0.85, where c is the rate at which Bitcoin transactions arrive in the mempool. Stoepker et al.~\cite{DBLP:journals/sigmetrics/StoepkerGK21} show by means of a robustness analysis that the time to ruin of a CL model is near insensitive to small changes in the model assumptions and illustrate that the proposed heuristic model can be used to accurately predict the confirmation times even when the data deviate (to a small degree) from the model assumptions. Ko et al.~\cite{DBLP:conf/blocksys/KoJMLH19} and Zhang et al.~\cite{DBLP:conf/wise/ZhangZLXL21} apply machine learning to the transactions data collected from the Bitcoin network based on a variety of factors including the historical confirmation time of transactions, block states, and mempool states.

\noindent
\emph{Analysis of mempool transactions}. At first, researchers focused on mempools because it was found that they could withstand DDos attacks by optimizing the mempool. Then the researchers focus on the transactions temporarily stored in the mempool.

Saad et al.~\cite{DBLP:conf/ccs/SaadTM18, DBLP:conf/icbc2/SaadNKKNM19} study the DDos attack on Bitcoin mempools and explore its effects on the mempool size and the transaction fees paid by users. Their countermeasures include fee-based and age-based designs, which optimize the mempool size and help to counter the effects of DDoS attacks.

Dae-Yong et al.~\cite{DBLP:conf/apnoms/KimEJ20} examine the variation of transactions in mempools through the Jaccard similarity index and analyze the causes of mempool difference. The result of the Jaccard index and Jaccard distance analysis show that most mempools are similar. Despite that, the mempool transactions are significantly different when a new block is produced. Kallurkar et al.~\cite{DBLP:conf/pkia/KallurkarC22} focus on statistics of failed cryptocurrency transactions and some primary reasons for failure in a cryptocurrency transaction. Furthermore, they also present existing approaches to minimize the failure of transactions.

The analysis of unconfirmed transactions in the mempool is still in its infancy, and the available research results are relatively few. To further explore the impact of the mempool on users, we focus on the privacy issue of unconfirmed transactions in the mempool, design novel clustering heuristics for unconfirmed transactions, and analyze the potential intentions of users from failed transactions.

\section{Conclusion and Future Work}
\label{sec:conclusion}

In this paper, we employed a combination of confirmed and unconfirmed transactions to cluster Bitcoin addresses, and demonstrated the influence of unconfirmed transactions on address clustering. The research suggests that unconfirmed transactions in the mempool can reduce the anonymity of Bitcoin, but it ultimately benefits Bitcoin developers and users by highlighting the issue and motivating further research into privacy protocols for the mempool. Individuals who are worried about safeguarding their privacy may opt to switch to cryptocurrencies that prioritize privacy, such as Zcash and Monero. However, prior investigations have demonstrated that even these currencies do not guarantee complete anonymity~\cite{DBLP:conf/uss/KapposYMM18, DBLP:conf/esorics/KumarFTS17}.


In the future, we aim to extend our analysis to other cryptocurrencies based on the UTXO model, such as Litecoin and Dogecoin. We will also analyze the Ethereum mempool and use the mempool data to explore the traceability of funds under the account-balance model. 


\balance
\small
\bibliographystyle{plain}
\bibliography{sample-base.bib}

\begin{thebibliography}{10}

\bibitem{DBLP:journals/corr/abs-1906-07852}
Cuneyt~Gurcan Akcora, Yitao Li, Yulia~R. Gel, and Murat Kantarcioglu.
\newblock Bitcoinheist: Topological data analysis for ransomware prediction on
  the bitcoin blockchain.
\newblock In {\em Proceedings of the 29th International Joint Conference on
  Artificial Intelligence (IJCAI)}, pages 4439--4445. ijcai.org, 2020.

\bibitem{alarab2020comparative}
Ismail Alarab, Simant Prakoonwit, and Mohamed~Ikbal Nacer.
\newblock Comparative analysis using supervised learning methods for anti-money
  laundering in bitcoin.
\newblock In {\em Proceedings of the 2020 5th International Conference on
  Machine Learning Technologies}, pages 11--17, 2020.

\bibitem{DBLP:conf/fc/AndroulakiKRSC13}
Elli Androulaki, Ghassan Karame, Marc Roeschlin, Tobias Scherer, and Srdjan
  Capkun.
\newblock Evaluating user privacy in bitcoin.
\newblock In {\em Proceedings of the 17th International Conference on Financial
  Cryptography and Data Security~(FC)}, volume 7859, pages 34--51. Springer,
  2013.

\bibitem{blockchain.com}
Inc. Blockchain.com.
\newblock The world’s most popular way to buy, sell, and trade crypto.
\newblock \url{https://www.blockchain.com/}, 2022.

\bibitem{DBLP:conf/complexnetworks/RemyRM17}
R{\'{e}}my Cazabet, Rym Baccour, and Matthieu Latapy.
\newblock Tracking bitcoin users activity using community detection on a
  network of weak signals.
\newblock In {\em Proceedings of the 6th International Conference on Complex
  Networks and Their Applications~(Complex Networks)}, pages 166--177.
  Springer, 2017.

\bibitem{DBLP:conf/icmla/ErmilovPY17}
Dmitry Ermilov, Maxim Panov, and Yury Yanovich.
\newblock Automatic bitcoin address clustering.
\newblock In {\em Proceedings of 16th International Conference on Machine
  Learning and Applications~(ICMLA)}, pages 461--466. {IEEE}, 2017.

\bibitem{foley2019sex}
Sean Foley, Jonathan~R Karlsen, and T{\=a}lis~J Putni{\c{n}}{\v{s}}.
\newblock Sex, drugs, and bitcoin: How much illegal activity is financed
  through cryptocurrencies?
\newblock {\em The Review of Financial Studies}, 32(5):1798--1853, 2019.

\bibitem{DBLP:journals/popets/GoldfederKRN18}
Steven Goldfeder, Harry~A. Kalodner, Dillon Reisman, and Arvind Narayanan.
\newblock When the cookie meets the blockchain: Privacy risks of web payments
  via cryptocurrencies.
\newblock {\em Proceedings on Privacy Enhancing Technologies},
  2018(4):179--199, 2018.

\bibitem{DBLP:journals/sigmetrics/GundlachGKR21}
Rowel G{\"{u}}ndlach, Martijn Gijsbers, David~T. Koops, and Jacques Resing.
\newblock Predicting confirmation times of bitcoin transactions.
\newblock {\em ACM SIGMETRICS Performance Evaluation Review}, 48(4):16--19,
  2021.

\bibitem{DBLP:conf/sp/HuangALIBMLLSM18}
Danny~Yuxing Huang, Maxwell~Matthaios Aliapoulios, Vector~Guo Li, Luca
  Invernizzi, Elie Bursztein, Kylie McRoberts, Jonathan Levin, Kirill
  Levchenko, Alex~C. Snoeren, and Damon McCoy.
\newblock Tracking ransomware end-to-end.
\newblock In {\em Proceedings of the 39th {IEEE} Symposium on Security and
  Privacy~(S\&P)}, pages 618--631. {IEEE}, 2018.

\bibitem{Chainalysis}
Chainalysis Inc.
\newblock Chainalysis: The blockchain data platform.
\newblock \url{https://www.chainalysis.com/}, 2022.

\bibitem{PingCAP}
PingCAP Inc.
\newblock Tidb: Open, unified, distributed sql.
\newblock \url{https://www.pingcap.com/}, 2022.

\bibitem{walletexplorer}
Aleš Janda.
\newblock Bitcoin block explorer with address grouping and wallet labeling.
\newblock \url{https://www.walletexplorer.com/}, 2022.

\bibitem{DBLP:conf/pkia/KallurkarC22}
Harshal~Shridhar Kallurkar and B.~R. Chandavarkar.
\newblock Unconfirmed transactions in cryptocurrency: Reasons, statistics, and
  mitigation.
\newblock In {\em Proceedings of 2022 International Conference on Public Key
  Infrastructure and its Applications~(PKIA)}, pages 1--7. {IEEE}, 2022.

\bibitem{DBLP:journals/corr/abs-1709-02489}
Harry~A. Kalodner, Malte M{\"{o}}ser, Kevin Lee, Steven Goldfeder, Martin
  Plattner, Alishah Chator, and Arvind Narayanan.
\newblock Blocksci: Design and applications of a blockchain analysis platform.
\newblock In {\em Proceedings of the 29th {USENIX} Security Symposium~(USENIX
  Security)}, pages 2721--2738. {USENIX} Association, 2020.

\bibitem{DBLP:conf/uss/KapposYMM18}
George Kappos, Haaroon Yousaf, Mary Maller, and Sarah Meiklejohn.
\newblock An empirical analysis of anonymity in zcash.
\newblock In {\em Proceedings of the 27th {USENIX} Security Symposium~(USENIX
  Security)}, pages 463--477. {USENIX} Association, 2018.

\bibitem{DBLP:conf/uss/KapposYSRHM22}
George Kappos, Haaroon Yousaf, Rainer St{\"{u}}tz, Sofia Rollet, Bernhard
  Haslhofer, and Sarah Meiklejohn.
\newblock How to peel a million: Validating and expanding bitcoin clusters.
\newblock In {\em Proceedings of the 31th USENIX Security Symposium (USENIX
  security)}, pages 2207--2223. {USENIX} Association, 2022.

\bibitem{DBLP:conf/apnoms/KimEJ20}
Dae{-}Yong Kim, Meryam Essaid, and Hongtaek Ju.
\newblock Examining bitcoin mempools resemblance using jaccard similarity
  index.
\newblock In {\em Proceedings of the 21st Asia-Pacific Network Operations and
  Management Symposium~(APNOMS)}, pages 287--290. {IEEE}, 2020.

\bibitem{DBLP:conf/blocksys/KoJMLH19}
Kyungchan Ko, Taeyeol Jeong, Sajan Maharjan, Chaehyeon Lee, and James~Won{-}Ki
  Hong.
\newblock Prediction of bitcoin transactions included in the next block.
\newblock In {\em Proceedings of the 1st International Conference on Blockchain
  and Trustworthy Systems~(BlockSys)}, pages 591--597. Springer, 2019.

\bibitem{DBLP:conf/esorics/KumarFTS17}
Amrit Kumar, Cl{\'{e}}ment Fischer, Shruti Tople, and Prateek Saxena.
\newblock A traceability analysis of monero's blockchain.
\newblock In {\em Proceedings of the 22nd European Symposium on Research in
  Computer Security~(ESORICS)}, pages 153--173. Springer, 2017.

\bibitem{DBLP:conf/blocksys/LiLZ19}
Yang Li, Zilu Liu, and Zibin Zheng.
\newblock Quantitative analysis of bitcoin transferred in bitcoin exchange.
\newblock In {\em Proceedings of the 1st International Conference on Blockchain
  and Trustworthy Systems~(BlockSys)}, pages 549--562. Springer, 2019.

\bibitem{DBLP:conf/iscisc/ManaviH20}
Farnoush Manavi and Ali Hamzeh.
\newblock A new method for ransomware detection based on {PE} header using
  convolutional neural networks.
\newblock In {\em Proceedings of the 17th International Conference on
  Information Security and Cryptology~(ISCISC)}, pages 82--87. {IEEE}, 2020.

\bibitem{DBLP:conf/iconip/McIntoshJWS19}
Timothy~R. McIntosh, Julian Jang-Jaccard, Paul~A. Watters, and Teo Susnjak.
\newblock The inadequacy of entropy-based ransomware detection.
\newblock In {\em Proceedings of the 26th International Conference on Neural
  Information Processing~(ICONIP)}, pages 181--189. Springer, 2019.

\bibitem{DBLP:conf/raid/MehnazMB18}
Shagufta Mehnaz, Anand Mudgerikar, and Elisa Bertino.
\newblock Rwguard: {A} real-time detection system against cryptographic
  ransomware.
\newblock In {\em Proceedings of the 21st International Symposium on Research
  in Attacks, Intrusions, and Defenses~(RAID)}, pages 114--136. Springer, 2018.

\bibitem{DBLP:conf/imc/MeiklejohnPJLMVS13}
Sarah Meiklejohn, Marjori Pomarole, Grant Jordan, Kirill Levchenko, Damon
  McCoy, Geoffrey~M. Voelker, and Stefan Savage.
\newblock A fistful of bitcoins: characterizing payments among men with no
  names.
\newblock In {\em Proceedings of the 2013 Internet Measurement
  Conference~(IMC)}, pages 127--140. {ACM}, 2013.

\bibitem{DBLP:journals/popets/Moreno-SanchezZ16}
Pedro Moreno{-}Sanchez, Muhammad~Bilal Zafar, and Aniket Kate.
\newblock Listening to whispers of ripple: Linking wallets and deanonymizing
  transactions in the ripple network.
\newblock {\em Proceedings on Privacy Enhancing Technologies},
  2016(4):436--453, 2016.

\bibitem{DBLP:conf/ecrime/MoserBB13}
Malte M{\"{o}}ser, Rainer B{\"{o}}hme, and Dominic Breuker.
\newblock An inquiry into money laundering tools in the bitcoin ecosystem.
\newblock In {\em Proceedings of the 2013 {APWG} eCrime Researchers
  Summit~(eCrime)}, pages 1--14. IEEE, 2013.

\bibitem{BTCWhitePaper}
Satoshi Nakamoto.
\newblock Bitcoin: A peer-to-peer electronic cash system, 2009.

\bibitem{DBLP:journals/cybersecurity/Paquet-Clouston19}
Masarah Paquet{-}Clouston, Bernhard Haslhofer, and Benoit Dupont.
\newblock Ransomware payments in the bitcoin ecosystem.
\newblock {\em Journal of Cybersecurity}, 5(1):tyz003, 2019.

\bibitem{DBLP:journals/fgcs/PierroRDMT22}
Giuseppe~Antonio Pierro, Henrique Rocha, St{\'{e}}phane Ducasse, Michele
  Marchesi, and Roberto Tonelli.
\newblock A user-oriented model for oracles' gas price prediction.
\newblock {\em Future Generation Computer Systems}, 128:142--157, 2022.

\bibitem{DBLP:conf/socialcom/ReidH11}
Fergal Reid and Martin Harrigan.
\newblock An analysis of anonymity in the bitcoin system.
\newblock In {\em Proceedings of the 3rd International Conference on Social
  Computing (SocialCom)}, pages 1318--1326. IEEE, 2011.

\bibitem{Reid2012An}
Fergal Reid and Martin Harrigan.
\newblock An analysis of anonymity in the bitcoin system.
\newblock In {\em Proceedings of the 2nd International Conference on Privacy,
  Security, Risk and Trust~(PASSAT)}, pages 1318--1326. Springer, 2011.

\bibitem{DBLP:conf/icbc2/SaadNKKNM19}
Muhammad Saad, Laurent Njilla, Charles~A. Kamhoua, Joongheon Kim, DaeHun Nyang,
  and Aziz Mohaisen.
\newblock Mempool optimization for defending against ddos attacks in pow-based
  blockchain systems.
\newblock In {\em Proceedings of the 2019 International Conference on
  Blockchain and Cryptocurrency~(ICBC)}, pages 285--292. {IEEE}, 2019.

\bibitem{DBLP:conf/ccs/SaadTM18}
Muhammad Saad, My~T. Thai, and Aziz Mohaisen.
\newblock {POSTER:} deterring ddos attacks on blockchain-based cryptocurrencies
  through mempool optimization.
\newblock In {\em Proceedings of the 2018 on Asia Conference on Computer and
  Communications Security~(AsiaCCS)}, pages 809--811. {ACM}, 2018.

\bibitem{BitcoinTool}
Michele Spagnuolo, Federico Maggi, and Stefano Zanero.
\newblock Bitiodine: Extracting intelligence from the bitcoin network.
\newblock In {\em Proceedings of the 18th International Conference on Financial
  Cryptography and Data Security~(FC)}, pages 457--468. Springer, 2014.

\bibitem{DBLP:journals/sigmetrics/StoepkerGK21}
Ivo Stoepker, Rowel G{\"{u}}ndlach, and Stella Kapodistria.
\newblock Robustness analysis of bitcoin confirmation times.
\newblock {\em ACM SIGMETRICS Performance Evaluation Review}, 48(4):20--23,
  2021.

\bibitem{DBLP:journals/tweb/WangPCZHCH22}
Kai Wang, Jun Pang, Dingjie Chen, Yu~Zhao, Dapeng Huang, Chen Chen, and Weili
  Han.
\newblock A large-scale empirical analysis of ransomware activities in bitcoin.
\newblock {\em ACM Transactions on the Web}, 16(2):7:1--7:29, 2022.

\bibitem{wang2018anti}
Yunpeng Wang, Jin Yang, Tao Li, Fangdong Zhu, and Xiaojun Zhou.
\newblock Anti-dust: A method for identifying and preventing blockchain’s
  dust attacks.
\newblock In {\em Proceedings of the 2018 International Conference on
  Information Systems and Computer Aided Education (ICISCAE)}, pages 274--280.
  IEEE, 2018.

\bibitem{DBLP:conf/wise/ZhangZLXL21}
Limeng Zhang, Rui Zhou, Qing Liu, Jiajie Xu, and Chengfei Liu.
\newblock Transaction confirmation time estimation in the bitcoin blockchain.
\newblock In {\em Proceedings of the 22nd International Conference on Web
  Information Systems Engineering~(WISE)}, pages 30--45. Springer, 2021.

\bibitem{DBLP:conf/wise/ZhangZLXL22}
Limeng Zhang, Rui Zhou, Qing Liu, Jiajie Xu, and Chengfei Liu.
\newblock Bitcoin transaction confirmation time prediction: {A} classification
  view.
\newblock In {\em Proceedings of the 23rd International Conference on Web
  Information Systems Engineering~(WISE)}, pages 155--169. Springer, 2022.

\bibitem{DBLP:journals/access/Zhang0020}
Yuhang Zhang, Jun Wang, and Jie Luo.
\newblock Heuristic-based address clustering in bitcoin.
\newblock {\em {IEEE} Access}, 8:210582--210591, 2020.

\bibitem{DBLP:journals/apin/ZolaSBGU22}
Francesco Zola, Lander Segurola{-}Gil, Jan~L. Bruse, Mikel Galar, and
  Raul~Orduna Urrutia.
\newblock Attacking bitcoin anonymity: generative adversarial networks for
  improving bitcoin entity classification.
\newblock {\em Applied Intelligence}, 52(15):17289--17314, 2022.

\end{thebibliography}


\end{document}